\documentclass[journal]{IEEEtran}
\usepackage{cite}
\usepackage{amsmath,amssymb,amsfonts}
\usepackage{algorithmic}
\usepackage{graphicx}
\usepackage{subfigure}
\usepackage{textcomp}
\usepackage{booktabs}
\usepackage{epstopdf}
\usepackage{array}
\usepackage{pgffor}
\usepackage{bm}
\usepackage{algorithm}
\usepackage{algorithmic}

\def\BibTeX{{\rm B\kern-.05em{\sc i\kern-.025em b}\kern-.08em
		T\kern-.1667em\lower.7ex\hbox{E}\kern-.125emX}}
\begin{document}
	
	\title{Analysis of Degrees of Freedom in Scattered Fields for Nonlinear Inverse Scattering Problems}
	
	\author{Zhichao Lin,
		Rui Guo, \IEEEmembership{Member, IEEE,}
		Tao Shan, 
		Maokun Li, \IEEEmembership{Senior Member, IEEE,} 
		Fan Yang, \IEEEmembership{Fellow, IEEE,}
		Shenheng Xu, \IEEEmembership{Member, IEEE,}
		and Aria Abubakar, \IEEEmembership{Senior Member, IEEE}
		
		\thanks{This work was supported in part by the National Natural Science Foundation of China under Grant 61971263, in part by the National Key R\&D Program of China under Grant 2018YFC0603604, and in part by the Institute for Precision Medicine, Tsinghua University, Beijing, China. \textit{(Corresponding author: Maokun Li.)}}
		\thanks{Zhichao Lin, Rui Guo, Tao Shan, Maokun Li, Fan Yang, and Shenheng Xu are with Beijing National Research Center for Information Science and Technology (BNRist), Department of Electronic Engineering, Tsinghua University, Beijing 100084, China (e-mail: maokunli@tsinghua.edu.cn).}
		\thanks{Aria Abubakar is with Schlumberger, Houston, TX 77056 USA (e-mail: aabubakar@slb.com).}
	}
	
	\maketitle
	
	\begin{abstract}
		In the study of nonlinear inverse scattering problems (ISPs), the traditional method for estimating the number of degrees of freedom (NDoF) of the scattered field does not consider the effect of the scatterer contrast. In this work, we study the relationship between NDoF and scatterer contrast by modifying the radiation matrix. Specifically, three numerical methods for calculating the NDoF are presented. The first NDoF is defined as the number of principle singular values of the external radiation matrix, which is used as a benchmark. The second NDoF is defined based on the modified radiation matrix that incorporates the contrast of domain of investigation (DoI) with the external radiation matrix. By exploring the relationship between the second NDoF and contrast, an approximate upper bound of the second NDoF is obtained, that is, the third NDoF. The accuracy of the proposed NDoF is verified by interpolation experiments on both synthetic and experimental data using frequency-domain zero-padding method. Results show that the second NDoF is more accurate than traditional NDoF for high-contrast scatterers, and the third NDoF can be easily estimated by the traditional NDoF plus a positive constant, which provides a simple way to estimate the upper bound of NDoF. 
	\end{abstract}
	
	
	\section{Introduction}
	Electromagnetic inverse scattering problem (ISP) can be considered as a mathematical model for a wide range of real-world applications, such as remote sensing, biomedical imaging, geophysical measurement, and nondestructive evaluation \cite{chen2018computational, woodhouse2017introduction, 9325547, abubakar20082, salucci2016real}. ISP aims to reconstruct the material properties and spatial distributions of the unknown scatterers in the domain of investigation (DoI) from the measured scattered fields. In general, the DoI is sequentially illuminated by a series of known transmitters, and the scattered field for each illumination is measured by a number of receivers around the DoI. Theoretically, the scattered field is a continuous function of spatial position. However, in practice, the measurement must be carried out at finite discrete positions. Therefore an important question arises on the minimum number of receivers needed to sample the scattered field. If there are too few receivers, the measured data cannot accurately represent the continuous scattered field, which will introduce additional ill-posedness to the ISP. If there are too many receivers, the oversampled data will cause an unnecessary increase in the time and memory cost for solving the ISP, not to mention the extra cost in the measurement. To find a proper number of receivers, the degrees of freedom (DoF) of the scattered field has been important in ISP research and many researchers have contributed to this topic.
	
	The number of degrees of freedom (NDoF) generally describes the minimum number of independent parameters for a system, and the degrees of freedom (DoF) refer to those independent basis components. In ISP, the NDoF of scattered field can be defined as the minimum number of parameters needed to completely characterize the continuous scattered field within a given precision, and DoF are the corresponding basis functions. This notion was first studied by Bucci et al \cite{bucci1987spatial, bucci1989degrees, bucci1997electromagnetic, bucci1999improving}. In \cite{bucci1987spatial}, the bandlimited property of the scattered field in the far-field region is studied, and the spatial effective bandwidth of the field is evaluated. In \cite{bucci1989degrees}, the NDoF of scattered field is computed, which is equal to the Nyquist number. Besides, the uniform sampling expansion is proven to be a practically optimal algorithm for representing the scattered field. In \cite{bucci1997electromagnetic}, the available information in single view and multiview cases are evaluated. The NDoF of scattered field estimated from the singular value decomposition (SVD) and effective bandwidth approaches are summarized, which are both equal to twice the electric size in 2D single view case. The accuracy of NDoF is verified by numerical examples of scattered field interpolation. In \cite{bucci1999improving}, the NDoF of scattered field in the near-field region is exploited by the singular value decomposition of the radiation operator. The analytical expressions of singular values and functions for 2D case are given, which shows that the NDoF in the near-field region is slightly increased and is susceptible to the noise level. Based on the concept of NDoF, non-redundant representation and interpolation of radiated fields for different cases are studied \cite{bucci1991optimal, bucci1993interpolation, bucci1998representation}. In addition, the information content of Born scattered fields in ISP are studied in \cite{brancaccio1998information, pierri1999information}. 
	
	DoF has also been widely applied in analysis of various electromagnetic fields. The early investigations on the NDoF are carried out in optical imaging applications \cite{di1969degrees}. In \cite{piestun2000electromagnetic}, the NDoF is defined as the number of orthogonal modes that account for most of the total connection strength. For radiated fields, the concept of DoF are also widely studied \cite{bertero1989linear}. In \cite{pierri1998information}, the information contents of radiated fields in both Fresnel zone and near-field zone of the source domain are investigated. In \cite{pierri2020asymptotic, pierri2021ndf, leone2022dimension}, the NDoF and optimal sampling methods of near-zone field are investigated. In \cite{pierri2021svd} and \cite{pierri2021dimension}, the dimensions of phaseless field data in Fresnel zone and near-field zone are studied, respectively. In \cite{moretta2022optimal}, the optimal field samplings of arc sources for both far and near fields are addressed. It is noted that the above works all use the singular value decomposition (SVD) of radiation operators to analyze the NDoF. For communication system, the DoF analysis of wireless communication systems based on multiple-input multiple-output (MIMO) technique has attracted much attention \cite{migliore2020cares}. The relationships between information theory and electromagnetic theory are widely studied \cite{migliore2008electromagnetics, loyka2018information, gruber2008new, franceschetti2015information}. The NDoF of MIMO systems based on single scattering approximation is studied in \cite{poon2005degrees, hanlen2006wireless}, and that in multiple scattering environment is evaluated in \cite{xu2006electromagnetic, janaswamy2011degrees}. It is noted that the NDoF in \cite{xu2006electromagnetic} is also based on the singular values of the radiation operator. In addition, DoF is applied to analyze the resolution limits of multipath-assisted radar imaging \cite{dickey2003super, krishnan2011synthetic, solimene2014sar, mehrotra2020dof}. In short, DoF provides a theoretical basis for analyzing the performance of an electromagnetic system as well as setting the state parameters of system for different applications. 
	
	Based on the aforementioned works, we study the impact of scatterer contrast on the NDoF of scattered field in nonlinear ISP. The NDoF is defined as the number of field samples needed to represent the continuous field within a given precision. Firstly, by reviewing the step-like behavior of singular values, the first NDoF is defined as the number of singular values that contribute to a significant fraction $P$ of all the singular values of the external radiation matrix. The first NDoF is applied as a benchmark. Secondly, the radiation matrix is modified to incorporate the maximum contrast of DoI under the assumption that DoI is homogeneous. The second NDoF is defined based on the modified radiation matrix. Thirdly, by investigating the relationship between the second NDoF and the maximum contrast, we assume that the maximum contrast of DoI is very large when it is unavailable, then an approximate upper bound of the second NDoF is obtained, that is, the third NDoF. It is noted that the third NDoF is independent of contrast. To verify the accuracy of the proposed NDoF, frequency-domain zero-padding method \cite{lyons2010interpolate, lyons2010understanding} is applied to interpolate the field samples of NDoF into oversampled fields. Interpolation examples using both synthetic and experimental data validate the effectiveness of the proposed NDoF.
	
	This paper is organized as follows. In Section \ref{sec2}, the forward modeling for 2D scalar wave propagation and the related works are briefly reviewed. In Section \ref{sec3}, the numerical methods for calculating the NDoF are presented. In Section \ref{sec4}, the relationship between NDoF and the contrast of DoI is investigated, and the accuracy of the proposed NDoF is verified. Discussions and conclusions are given in Section \ref{sec5} and Section \ref{sec6}, respectively.
	
	\section{Formulation} \label{sec2}
	\begin{figure}[tbp]
		\centering
		\includegraphics[width=0.7\linewidth]{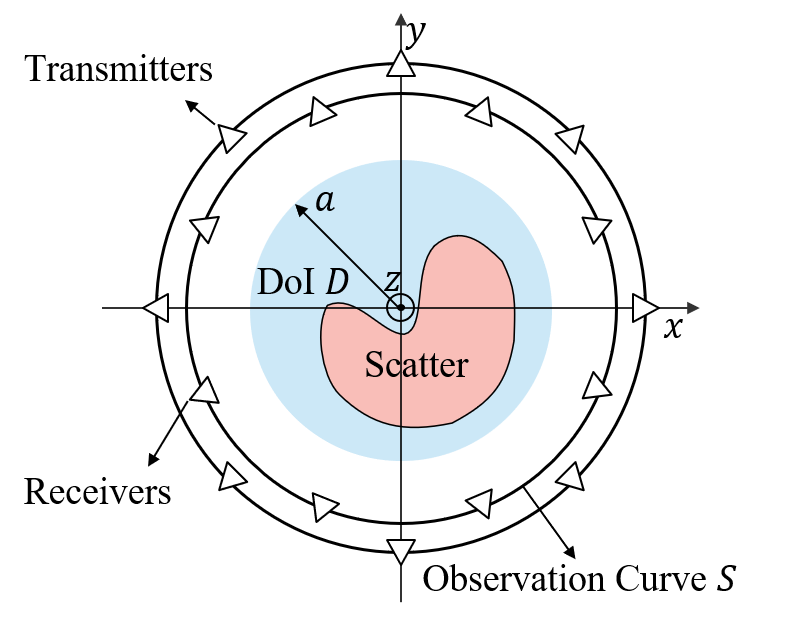}
		\caption{Geometry of 2D inverse scattering scenario.}
		\label{Fig1}
	\end{figure}
	We consider a 2D scenario with TM-polarized line sources. The background is a free space with permittivity $\varepsilon_0$, and the DoI $D$ is a circle of radius $a$ in cross section, as shown in Fig. \ref{Fig1}. The permittivity distribution of $D$ is $\varepsilon(\mathbf{r})$, where $\mathbf{r}$ is the 2D spatial location. Assume that transmitters and receivers are evenly located on the observation circle $S$ of radius $R_o$ around $D$. Without loss of generality, we set their numbers to be the same, denoted as $N_r$. Given the incident field $E^{inc}(\mathbf{r})$, the total electric field $E^{tot}(\mathbf{r})$ and the scattered field $E^{sca}(\mathbf{r})$ satisfy the following integral equations \cite{chen2018computational}
	\begin{align}
		E^{tot}(\mathbf{r})&=E^{inc}(\mathbf{r})+\int_{D} G\left(\mathbf{r}, \mathbf{r}^{\prime}\right) \chi\left(\mathbf{r}^{\prime}\right) E^{tot}\left(\mathbf{r}^{\prime}\right) d\mathbf{r}^{\prime}  \notag \\ 
		&=E^{inc}(\mathbf{r})+G_D\left(\chi E^{tot}\right), \quad \mathbf{r} \in D \label{eq1} \\
		E^{sca}(\mathbf{r})&=\int_{D} G\left(\mathbf{r}, \mathbf{r}^{\prime}\right) \chi\left(\mathbf{r}^{\prime}\right) E^{tot}\left(\mathbf{r}^{\prime}\right) d\mathbf{r}^{\prime}  \notag \\
		&=G_S\left(\chi E^{tot}\right), \quad \mathbf{r} \in S \label{eq2}
	\end{align}
	where $G\left(\mathbf{r}, \mathbf{r}^{\prime}\right) = -(j\beta^2/4)H_0^{(2)}(\beta|\mathbf{r}-\mathbf{r}^{\prime}|)$ is the 2D free space Green's function, $\beta$ is the wavenumber of free space, $H_0^{(2)}$ is the zeroth-order Hankel function of the second kind, and $\chi(\mathbf{r})$ is the contrast function defined as $\chi(\mathbf{r}) = \varepsilon(\mathbf{r}) / \varepsilon_0 - 1$ that represents the electromagnetic properties of the unknown scatterers. $G_D$ and $G_S$ are the internal and external radiation operators, respectively.
	
	For the above model, the normalized singular values of the external radiation operator $G_S$ are given by \cite{bucci1999improving}
	\begin{equation} \label{eq3}
		\hat{\sigma}_{n}^{2} =\frac{\sigma_{n}^{2}}{\sigma_{0}^{2}} 
		=\left|\frac{H_{n}^{(2)}\left(\beta R_{o}\right)}{H_{0}^{(2)}\left(\beta R_{o}\right)}\right|^{2} \frac{J_{n}^{2}\left(\beta a\right)-J_{n-1}\left(\beta a\right) J_{n+1}\left(\beta a\right)}{J_{0}^{2}\left(\beta a\right)+J_{1}^{2}\left(\beta a\right)}
	\end{equation}
	where $J_n$ is $n$th-order Bessel function, $H_n^{(2)}$ is the $n$th-order Hankel function of the second kind, and the index $n$ ranges from $-\infty$ to $+\infty$. For each value of $|n|\neq0$, there is a pair of identical singular values with different singular functions. The first term of \eqref{eq3} shows the impact of the radius of the observation circle on the singular values, which tends to unity as $R_o\rightarrow \infty$. Thus, the second term represents the behavior of singular values in the far-field region.
	
	For electrically large scatterers ($\beta a \gg1$), the singular values in the far-field region exhibit a step-like behavior with a knee after the first $2\beta  a$ values. Then, the singular values decay exponentially, and the transition region decreases with the increase of electric size of scatterers \cite{bucci1997electromagnetic}. This property allows to define the NDoF of scattered field as $2\beta  a$ regardless of the prescribed error. However, although using singular functions to represent the scattered field has the minimum reconstruction error, it is not convenient. More specifically, the coefficients of each singular function are calculated based on the entire field. As an alternative, the uniform sampling expansion is proved to be a practically optimal basis for representing the scattered field, which leads to a marginal increase in the number of terms of the representation compared to the singular functions \cite{bucci1989degrees}. The interpolation formula for this 2D geometry is given by \cite{bucci1997electromagnetic}
	\begin{equation}
		\label{eq4}
		\begin{aligned}
			E^{sca}\left(\theta^o, \theta^i\right) =& \sum_{n, m=-N}^{N} E^{sca}\left(\theta_{n}^{o}, \theta_{m}^{i}\right)\cdot \\ &D_N\left(\theta^o-\theta^o_{n}\right)D_N\left(\theta^i-\theta^i_{m}\right)
		\end{aligned}
	\end{equation}
	where $\theta^o_n=2\pi n/(2N+1)$ is the position of $n$th receiver, $\theta^i_m=2\pi m/(2N+1)$ is the position of $m$th transmitter, and $D_N$ is the Dirichlet function of degree $N$ defined as
	\begin{equation}
		D_{N}(\theta)=\frac{\sin \left(\frac{2 N+1}{2} \theta\right)}{(2 N+1) \sin (\theta / 2)}
	\end{equation}
	It is noted that to use this formula, the number of receivers and transmitters should be odd.
	
	In the above analysis, the electromagnetic properties of the scatterers are taken into account in the equivalent sources $J=j\omega \varepsilon_0 \chi E^{tot}$, which is assumed to be bounded by \cite{bucci1989degrees}
	\begin{equation}
		\int_{D}\left|J\left(\mathbf{r}^{\prime}\right)\right| d \mathbf{r}^{\prime} \leq C a
	\end{equation}
	where $C$ is a constant. Therefore, the value of material property is not considered in this analysis although its value actually has an effect on the approximation error of the scattered field \cite{bucci1987spatial}.
	
	\section{The Number of Degrees of Freedom} \label{sec3}
	To take into account the impact of scatterer contrast on the NDoF of scattered field, we resort to numerical methods. The method of moments (MoM) \cite{jin2011theory} is applied to solve the integral equations \eqref{eq1} and \eqref{eq2}. The DoI $D$ is partitioned into $M$ discrete subdomains. Accordingly, \eqref{eq1} and \eqref{eq2} can be written as matrix forms:
	\begin{align}
		&\bar{E}^{tot} = \bar{E}^{inc} + \bar{\bar{G}}_D \cdot \bar{\bar{X}} \cdot \bar{E}^{tot} \label{eq7}\\
		&\bar{E}^{sca} = \bar{\bar{G}}_S \cdot  \bar{\bar{X}} \cdot \bar{E}^{tot} \label{eq8}
	\end{align}
	where $\bar{E}^{tot}, \bar{E}^{inc}$ are $M$-dimensional vectors representing the total field and incident field values in each subdomain, respectively; $\bar{E}^{sca}$ is an $N_r$-dimensional vector representing the scattered field values measured on each receiver, $\bar{\bar{X}}$ is a diagonal matrix composed of the contrasts of each subdomain. $\bar{\bar{G}}_D$ and $\bar{\bar{G}}_S$ are the matrix versions of internal and external radiation operators, respectively. They can be computed by
	\begin{align}
		&\bar{\bar{G}}_D(m, n)
		=\left\{\begin{array}{ll}
			c H_{1}^{(2)}\left(\beta r_{e q}\right)-1, & m=n \\
			c J_{1}\left(\beta r_{e q}\right) H_{0}^{(2)}\left(\beta\left|\bar{r}_{m}-\bar{r}_{n}\right|\right), & m \neq n
		\end{array}\right.  \label{eq5}\\
		&\bar{\bar{G}}_S(p, n) = c J_{1}\left(\beta r_{eq}\right) H_{0}^{(2)}\left(\beta\left|\bar{r}_{p}-\bar{r}_{n}\right|\right)
	\end{align}
	where $\bar{r}_m, \bar{r}_n (m, n=1,2,\cdots,M)$ are the location vectors of each subdomain, $\bar{r}_p (p = 1,2,\cdots,N_r)$ is that of the $p^{\text{th}}$ receiver, $c=-(j \pi \beta r_{e q})/2$ is a constant coefficient, and $r_{eq}=\sqrt{S/\pi}$ is the equivalent radius of the subdomain where $S$ is the area of the subdomain \cite{jin2011theory}. 
	
	The singular value decomposition (SVD) can be applied to the radiation matrix $\bar{\bar{G}}_S$, and \eqref{eq8} is rewritten as
	\begin{equation}
		\bar{E}^{sca} = \sum_{k=1}^{N_r} \sigma_k \bar{u}_k \bar{v}^H_k \bar{J}^{d} \label{eq11}
	\end{equation}
	where the superscript $^H$ indicates complex conjugate transpose, and $\bar{J}^{d} = \bar{\bar{X}} \cdot \bar{E}^{tot}$ is the displacement current vector. The symbols $\sigma$, $\bar{u}$ and $\bar{v}$ denote the singular values, the left singular vectors, and the right singular vectors of the external radiation matrix $\bar{\bar{G}}_S$, respectively. As mentioned above, if the singular values $\sigma_k$ are sorted in nonincreasing order, they exhibit a step-like behavior with exponential decay, thus any scattered field can be represented by finite expansion terms within a certain approximation error \cite{bucci1997electromagnetic}. It is noted that this property is also investigated and utilized by the well-known subspace-based optimization method (SOM) \cite{chen2009subspace, zhong2009twofold}. Accordingly, the NDoF of scattered field can be defined by the number of the first few principal singular values,
	\begin{equation}
		N_{\text{DoF}}^{(1)} = \min\left\{N:\sum_{n=1}^{N}\sigma_n^{(1)}>P\sum_{n=1}^{N_r}\sigma_n^{(1)}\right\} \label{eq12}
	\end{equation}
	where $\sigma_1^{(1)}>\sigma_2^{(1)}>\cdots$ are the singular values of $\bar{\bar{G}}_S$ sorted in nonincreasing order, the superscript $^{(1)}$ implies the first method for calculating NDoF, $P$ is a predefined fraction that controls the approximation error of scattered field. For simplicity, we consider $P = 99\%$ as a typical value. More details about selecting a proper $P$ are given in Section \ref{sec5}.
	
	The scattered field vector $\bar{E}^{sca}$ can be represented as a span of the left singular vectors, where the coefficient of each vector $\bar{u}_k$ is $\sigma_k\bar{v}_k^H\bar{J}^d$. Therefore, though $\|\bar{J}^{d}\|$ is bounded, it affects each coefficient. $\bar{J}^d$ is nonlinear with the contrast distribution $\bar{\bar{X}}$. To study the relationship between contrast and NDoF, we calculate $\bar{J}^d$ from \eqref{eq7} and substitute it into \eqref{eq8}, 
	\begin{equation}
		\bar{E}^{sca} = \bar{\bar{G}}_S \left( \bar{\bar{I}} - \bar{\bar{X}} \cdot \bar{\bar{G}}_D \right)^{-1} \bar{\bar{X}} \cdot \bar{E}^{inc} \label{eq13}
	\end{equation}
	where $\bar{\bar{I}}$ is an identity matrix. For simplicity and generality, we assume that the DoI is homogeneous with the contrast equals to the maximum value of the contrast distribution $\bar{\bar{X}}$, i.e., $\chi=\max \bar{\bar{X}}$, then we can rewrite \eqref{eq13} as
	\begin{align}
		&\bar{E}^{sca} = \bar{\bar{G}}_S^\prime \chi \bar{E}^{inc}\\
		&\bar{\bar{G}}_S^{\prime} = \bar{\bar{G}}_S\left(\bar{\bar{I}}-\chi\bar{\bar{G}}_D\right)^{-1} \label{eq15}
	\end{align}
	According to \eqref{eq15}, the modified NDoF can be defined as
	\begin{equation}
		N_{\text{DoF}}^{(2)} = \min\left\{N:\sum_{n=1}^{N}\sigma_n^{(2)}>P\sum_{n=1}^{N_r}\sigma_n^{(2)}\right\} \label{eq16}
	\end{equation}
	where $\sigma_n^{(2)}$ is the singular values of $\bar{\bar{G}}_S^{\prime}$. In this way, the maximum value of contrast in DoI is considered in the calculation of NDoF.
	
	We observe from numerical experiments that the spatial bandwidth \cite{bucci1987spatial} of the scattered field first increases with the contrast of scatterers before it saturates, and then fluctuates around the saturation value. The higher the operating frequency, the faster it will saturate. A similar relationship is also exhibited between $N_{\text{DoF}}^{(2)}$ and the maximum contrast of DoI $\chi$. Thus, when $\chi$ is not available, we can assume that $\chi$ is large enough that the identity matrix $\bar{\bar{I}}$ in \eqref{eq13} can be ignored, and an approximate upper bound of NDoF is obtained as follows,
	\begin{align}
		&\bar{E}^{sca} = -\chi^{-1}\bar{\bar{G}}_S^{\prime\prime} \chi \bar{E}^{inc}\\
		&\bar{\bar{G}}_S^{\prime\prime} = \bar{\bar{G}}_S \bar{\bar{G}}_D^{-1} \\
		&N_{\text{DoF}}^{(3)} = \min\left\{N:\sum_{n=1}^{N}\sigma_n^{(3)}>P\sum_{n=1}^{N_r}\sigma_n^{(3)}\right\} \label{eq18}
	\end{align}
	where $\sigma_n^{(3)}$ is the singular values of $\bar{\bar{G}}_S^{\prime\prime}$. It is noted that $N_{\text{DoF}}^{(3)}$ is independent of the contrast $\chi$. In addition, according to the definitions of $N_{\text{DoF}}^{(1)}$, $N_{\text{DoF}}^{(2)}$, and $N_{\text{DoF}}^{(3)}$, it is easy to obtain that $N_{\text{DoF}}^{(1)} = N_{\text{DoF}}^{(2)}(\chi)$ when $\chi$ is 0, and  $N_{\text{DoF}}^{(3)} \approx N_{\text{DoF}}^{(2)}(\chi)$ when $\chi$ is very large.
	
	The accuracy of the proposed NDoF is verified by reconstructing the oversampled field from $N_{\text{DoF}}$ field samples, and the uniform sampling strategy is applied to sample the measurement data with the obtained NDoF, similar to \cite{bucci1997electromagnetic}. Specifically, for the case illustrated in Fig. \ref{Fig1}, we set up the transmitters and receivers evenly on the observation circle, and apply the frequency-domain zero-padding method \cite{lyons2010interpolate, lyons2010understanding} to interpolate the field. The interpolation process of $N_{\text{DoF}}$ field samples for one transmitter is shown in Algorithm \ref{algorithm1}. 
	\begin{algorithm}[h]
		\caption{Frequency-domain zero-padding method}
		\label{algorithm1}
		\begin{algorithmic}[1]
			\REQUIRE $\bar{x}:$ field samples, $m:$ length of target field
			\ENSURE $\bar{y}:$ interpolated field 
			\STATE $n = \text{length of}~\bar{x}$
			\STATE $\bar{f} = \text{DFT}(\bar{x})$ \% Discrete Fourier Transform
			\IF{$n$ is even}
			\STATE $q = n/2$
			\STATE $\bar{g} = [f_1, \cdots, f_q, \frac{f_{q+1}}{2}, 0,\cdots, 0, \frac{f_{q+1}}{2}, f_{q+2}, \cdots, f_n]$ \\ \% insert $(m-n-1)$ zeros
			\ELSE
			\STATE $q = (n+1)/2$
			\STATE $\bar{g} = [f_1, \cdots, f_q, 0,\cdots, 0, f_{q+1}, \cdots, f_n]$ \\ \% insert $(m-n)$ zeros
			\ENDIF
			\STATE $\bar{y} = m/n\times \text{IDFT}(\bar{g})$
		\end{algorithmic}
	\end{algorithm}
	
	For the scattered fields generated by $N_{\text{DoF}}$ transmitters, the 1D scattered field vector of size $N_{\text{DoF}} \times 1$ becomes a 2D matrix of size $N_{\text{DoF}} \times N_{\text{DoF}}$, where one dimension represents the receivers, and the other represents the transmitters. In this case, we can first interpolate the scattered field along the receiver dimension and then the transmitter dimension. It is worth noting that this method is equivalent to the optimal interpolation function \eqref{eq4} when $n$ is odd, where the former interpolates the field in frequency domain, and the latter interpolates the field in time domain. Besides, this method also removes the restriction that the number of field samples must be odd.
	
	\section{Numerical examples} \label{sec4}
	In this section, we validate the above analysis on NDoF with numerical experiments. The radius of DoI, denoted as $a$, is 0.1m, and that of observation circle, denoted as $R_o$, is 1.67m. The operating frequency is constrained to be no less than 2 GHz so that the DoI is in the far-field region of the transmitters. To discretize the circular DoI, we first define the domain of computation as the smallest square enclosing the circular DoI, and partition the domain into 128$\times$128 subdomains, and then discard the pixels outside the circular DoI in the calculation of NDoF. Therefore, the circular DoI is divided into 12892 subdomains. To calculate the NDoF through the proposed methods, the original $N_r$ must be larger than the NDoF. In the following examples, $N_r$ is set to 360, and the size of oversampled fields is 360$\times$360 if not indicated otherwise. The performance of interpolation is measured by the signal-to-noise ratio (SNR), which is defined by
	\begin{equation}
		\text{SNR} = 20 \log_{10} \frac{\|\bar{\bar{E}}_o\|_F}{\|\bar{\bar{E}}_o-\bar{\bar{E}}_i\|_F}
	\end{equation}
	where $\bar{\bar{E}}_o$ and $\bar{\bar{E}}_i$ are the oversampled and interpolated scattered fields, respectively; $\|\cdot\|_F$ represents the Frobenius norm of the matrix.
	
	\subsection{Homogeneous Dielectric Cylinder}
	\begin{figure}[!tbp]
	\centering
	\includegraphics[width=0.95\linewidth]{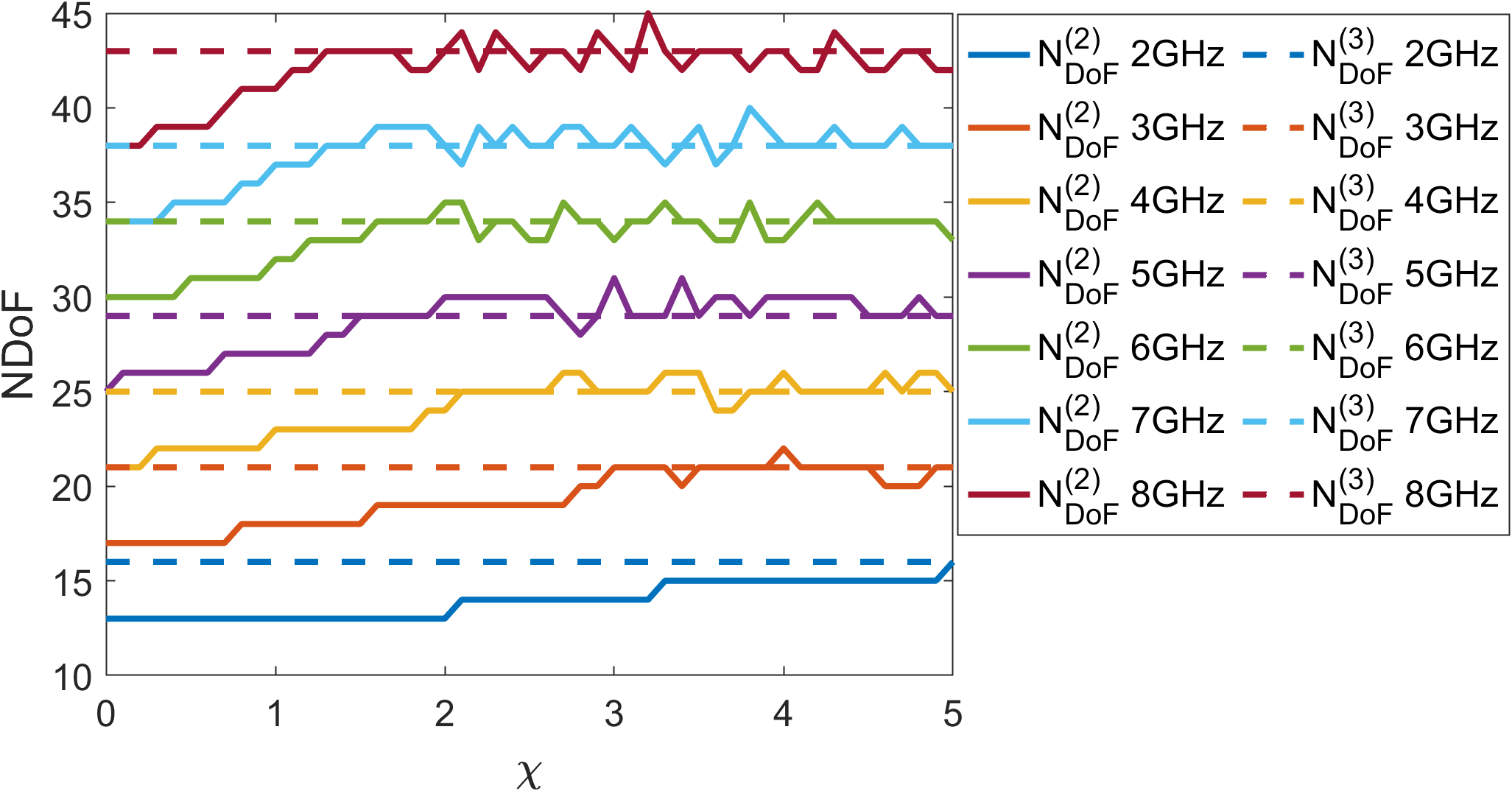}
	\caption{The NDoF versus the contrast of DoI. $N_{\text{DoF}}^{(1)}=N_{\text{DoF}}^{(2)}(\chi)$ when $\chi$ is 0, and $N_{\text{DoF}}^{(3)}\approx N_{\text{DoF}}^{(2)}(\chi)$ when $\chi$ is very large.}
	\label{Fig2}
	\end{figure}	
	According to the analytical analysis of DoF in Section \ref{sec2}, we first study the ideal case where the DoI is a homogeneous dielectric cylinder. To verify the relationship between NDoF and the contrast of DoI, we set the operating frequency ranging from 2 GHz to 8 GHz with the step of 1 GHz, and the material contrast $\chi$ varies from 0 to 5 with the step of 0.1. We compute NDoF using the three schemes discussed in the previous section, i.e., \eqref{eq12}, \eqref{eq16}, and \eqref{eq18}. The results are shown in Fig. \ref{Fig2}. It is noted that $N_{\text{DoF}}^{(1)}$ is equal to $N_{\text{DoF}}^{(2)}$ when $\chi$ is 0, and the approximate upper bound of $N_{\text{DoF}}^{(2)}$ of each frequency is illustrated in dash lines with the same color, i.e., $N_{\text{DoF}}^{(3)}$. From the figure, we can observe that $N_{\text{DoF}}^{(2)}$ first increases with the contrast before it reaches the upper bound $N_{\text{DoF}}^{(3)}$, and then fluctuates around the upper bound value. Moreover, the higher the operating frequency, the faster it reaches the upper bound. This behavior can be explained by the relationship between the degree of nonlinearity and the operating frequency, as analyzed in \cite{lin2021low}. Mathematically, the higher the frequency, the larger the norm of $G_D$, so according to \eqref{eq15}, the faster $N_{\text{DoF}}^{(2)}$ will be saturated. 	In fact, we notice that $N_{\text{DoF}}^{(3)}$ is not the upper bound of $N_{\text{DoF}}^{(2)}$ in the strict sense, but it basically estimates the maximum value of $N_{\text{DoF}}^{(2)}$, so we can regard $N_{\text{DoF}}^{(3)}$ as an approximate upper bound in practice. 
	
	\begin{figure}[!tbp]
		\centering
		\subfigure[]{
			\includegraphics[width=0.49\linewidth]{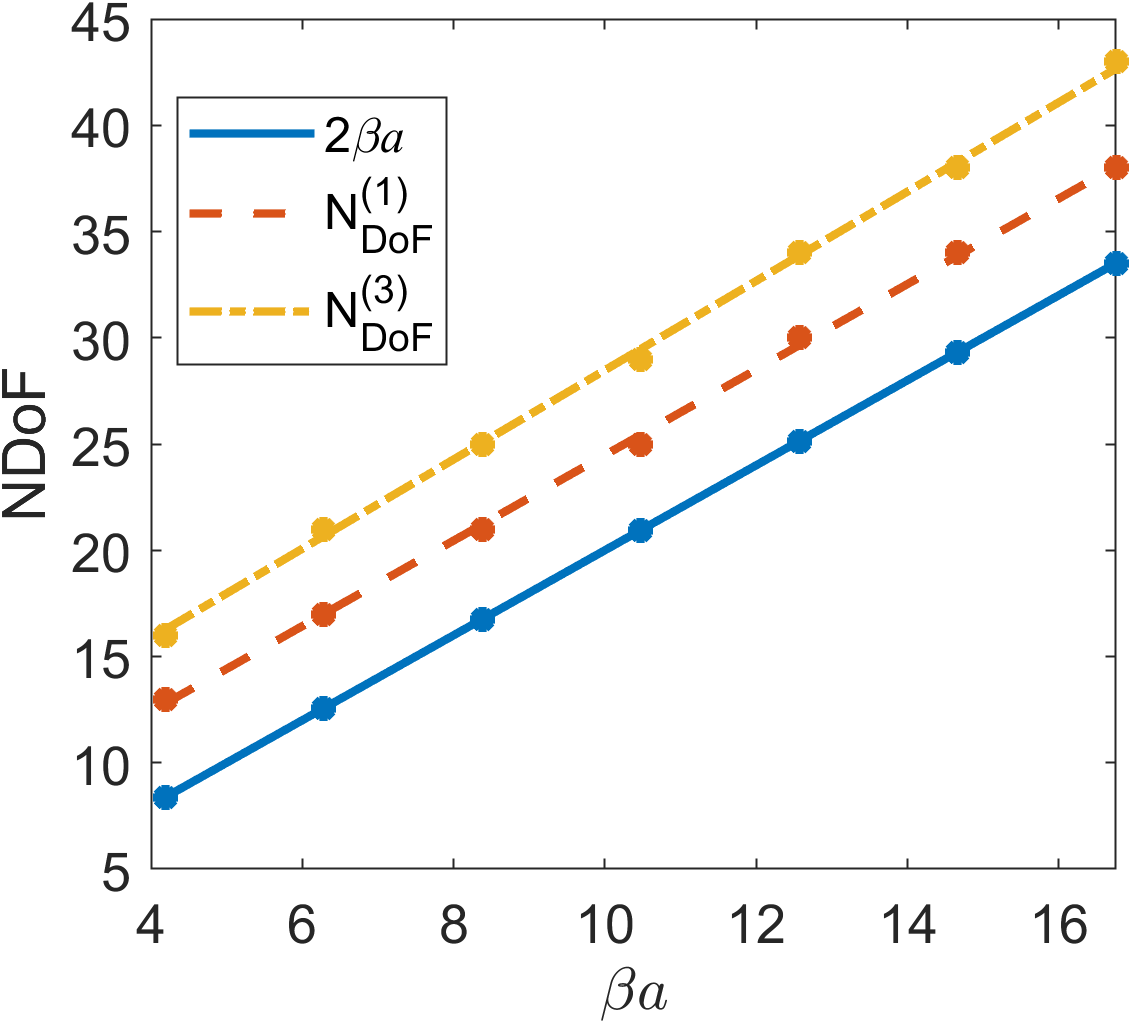}\label{Fig3a}}~
		\subfigure[]{
			\includegraphics[width=0.49\linewidth]{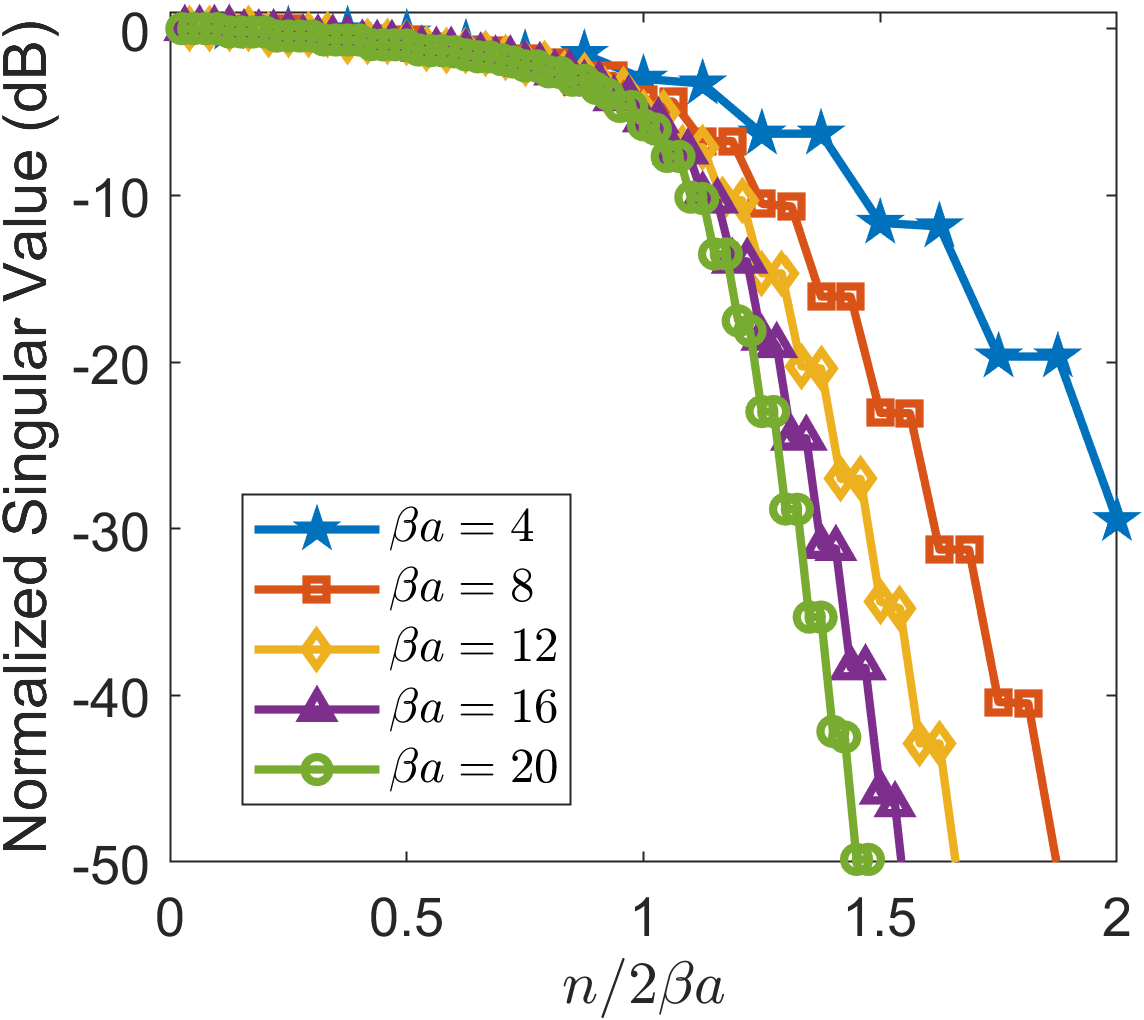}\label{Fig3b}}
		\caption{The relationship between NDoF and the electric size $\beta a$. (a) The NDoF versus the electric size $\beta a$. (b) Behavior of normalized singular values of $G_S^{\prime\prime}$ for different electric size $\beta a$. The index is normalized by $2\beta a$.}
		\label{Fig3}
	\end{figure}

	\begin{figure}[!tbp]
		\centering
		\includegraphics[width=0.98\linewidth]{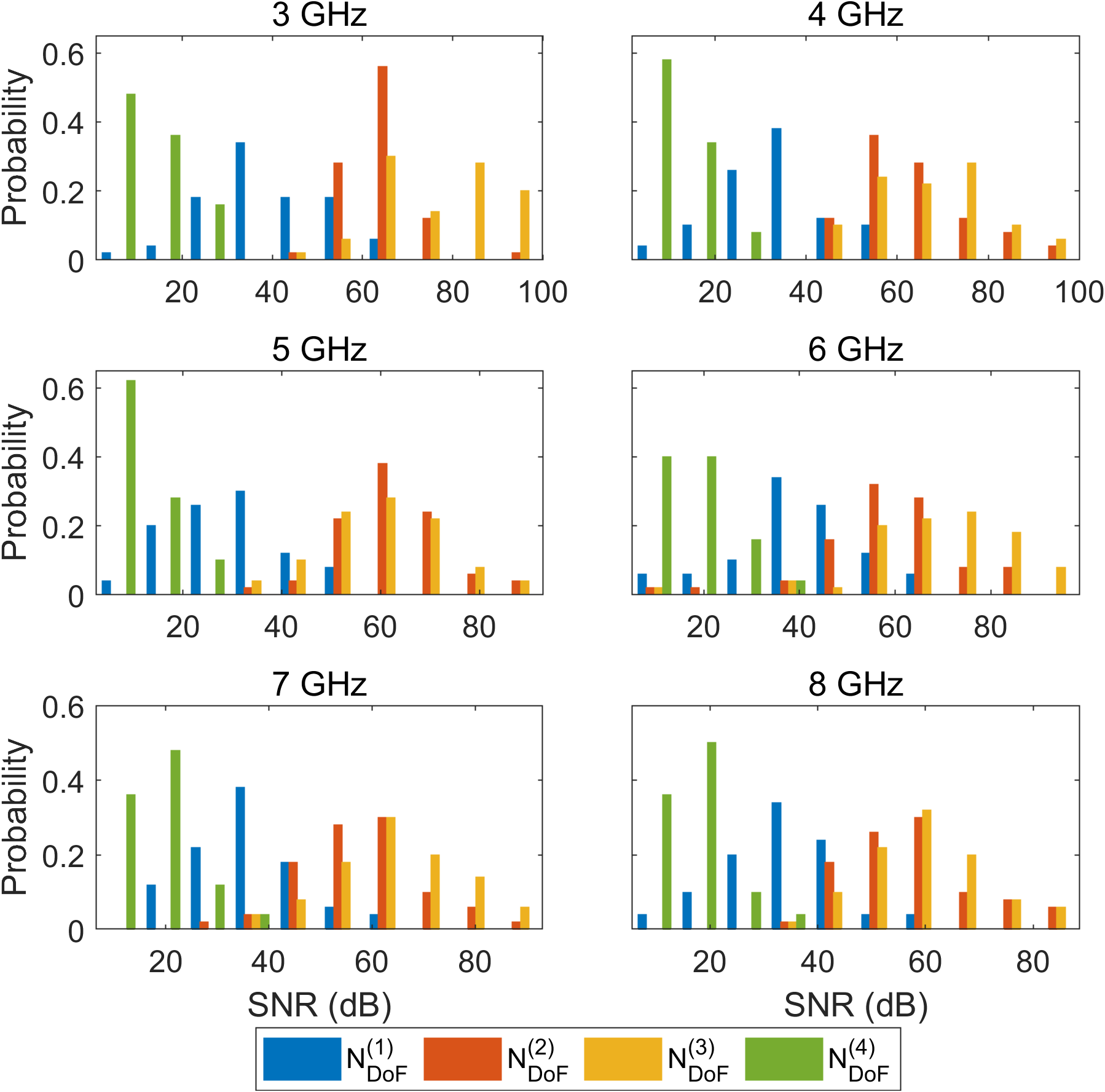}
		\caption{The SNR distributions of interpolation results for different operating frequencies. The contrast of DoI ranges from 0 to 5 with the step of 0.1.}
		\label{Fig4}
	\end{figure}
	
	Secondly, we study the relationship between NDoF and the electric size $\beta a$, as shown in Fig. \ref{Fig3}. The NDoF versus the electric size $\beta a$ is illustrated in Fig. \ref{Fig3a}. As mentioned in Section \ref{sec2}, previous works have indicated that the NDoF of scattered field is proportional to the electric size of the targets. More specifically, NDoF is approximately equal to $2\beta a$ for large scatterers, as shown by the blue line in Fig. \ref{Fig3a}. It can be seen that the red line and the yellow line are basically parallel to the blue line, where their slopes are 2.012 and 2.100, and biases are 4.36 and 7.47, respectively. Therefore, the proposed NDoF can be estimated by $2\beta a$ plus a proper constant bias for convenience. Besides, the behavior of normalized singular values of $G_S^{\prime\prime}$ for different electric sizes is shown in Fig. \ref{Fig3b}. It is obvious that the larger the electric size of targets, the faster the singular value decreases after the first $2\beta a$ principal values. Therefore, when the electric size is small, the NDoF will depend on the preset approximation error, which is consistent with the analysis presented in \cite{bucci1999improving}. 
	
	\begin{figure}[!tbp]
		\centering
		\includegraphics[width=0.9\linewidth]{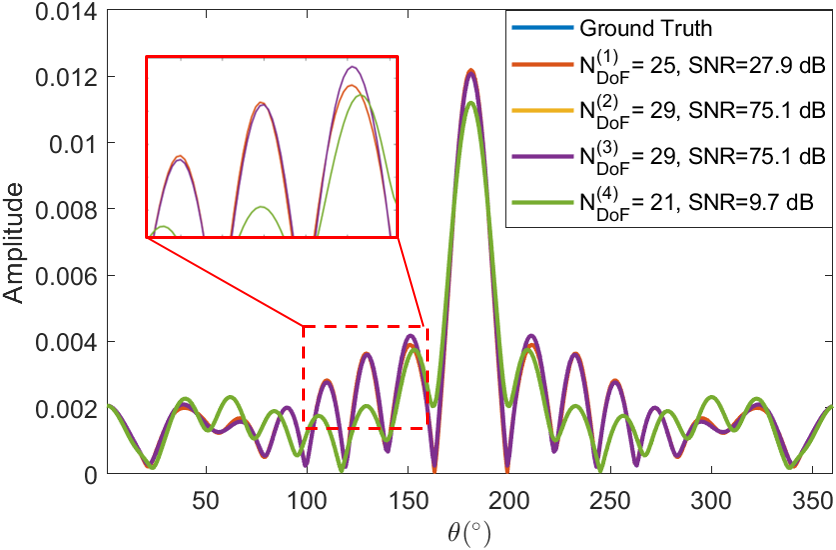}
		\caption{The interpolated scattered fields of four cases. The contrast of the cylinder is 5, and the operating frequency is 5 GHz.}
		\label{Fig5}
	\end{figure}

	\begin{figure}[!tbp]
	\centering
	\subfigure[]{
		\includegraphics[width=0.42\linewidth]{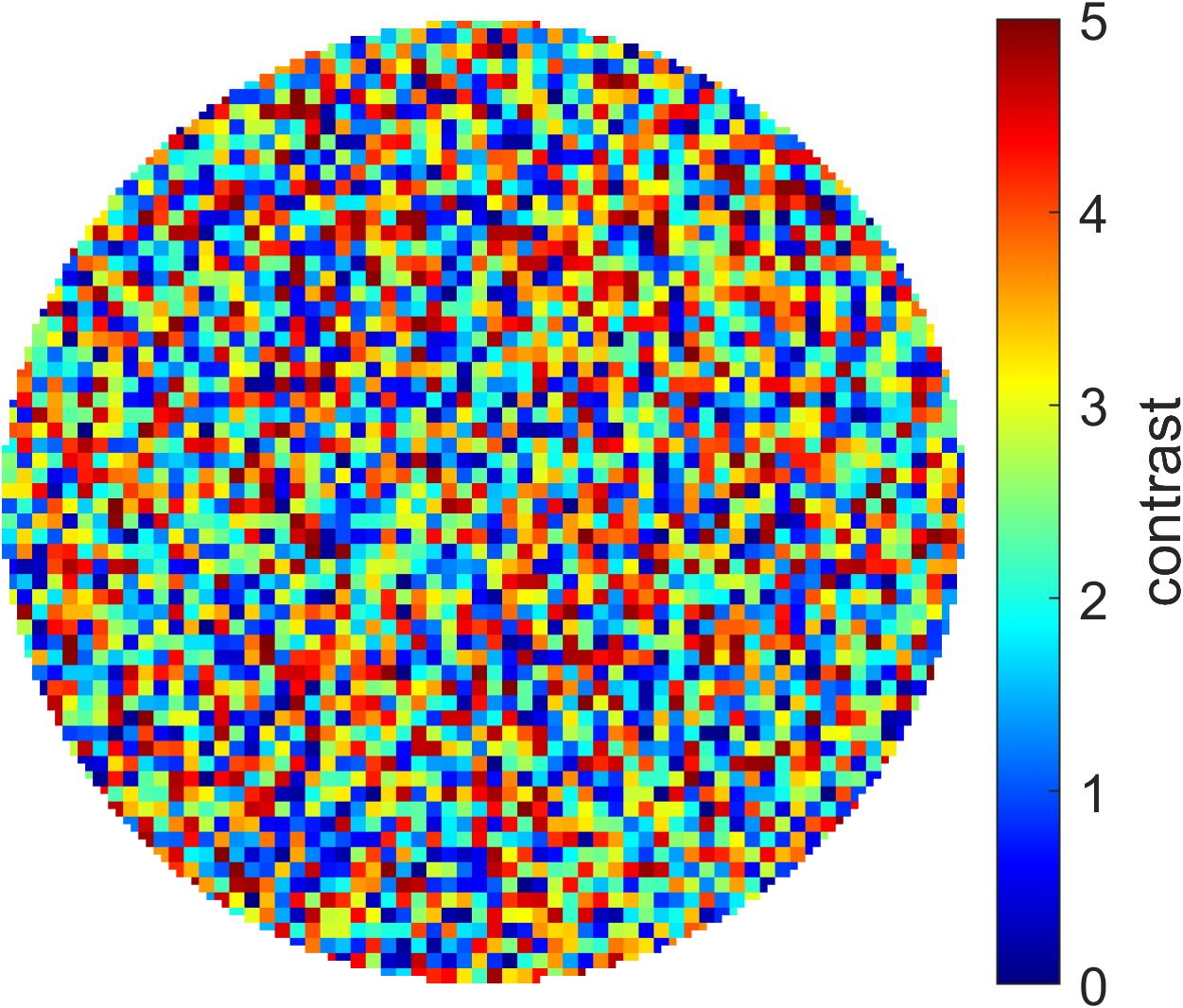}\label{Fig6a}}~~
	\subfigure[]{
		\includegraphics[width=0.42\linewidth]{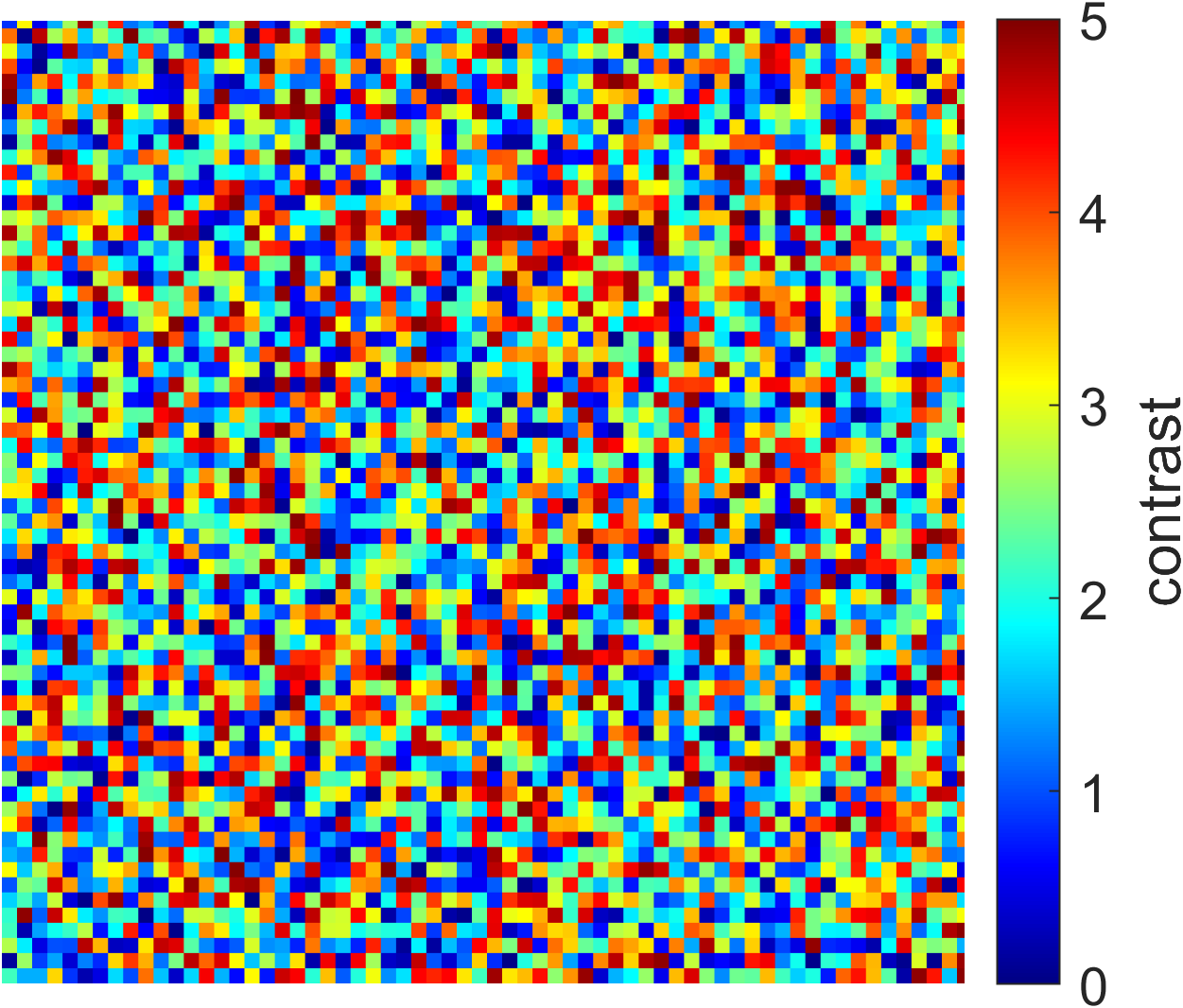}\label{Fig6b}}
	\caption{Schematic diagram of random dielectric samples. The contrast of each subdomain is randomly sampled from 0 to 5. (a) circular DoI. (b) square DoI.}
	\label{Fig6}
	\end{figure}
	
	Thirdly, we validate the proposed NDoF with interpolation experiments. The analytical NDoF $2\beta a$ is also considered. Since $2\beta a$ is not an integer, we define $N_{\text{DoF}}^{(4)}=\lceil 2\beta a \rceil$ as the smallest integer larger than $2\beta a$. To avoid numerical errors, the analytical solution of the scattered field of a homogeneous dielectric cylinder generated by a line source is applied in the experiment, which is given by \cite{jin2011theory},
	\begin{equation}
		\begin{aligned}
			E^{sca}(\rho,\phi)& = \frac{j}{4} \sum_{n=-\infty}^{+\infty} H_n^{(2)}(k\rho^{\prime})H_n^{(2)}(k\rho)e^{jn(\phi-\phi^{\prime})} \times \\
			&\frac{\sqrt{\mu_r}J_n^{\prime}(ka)J_n(k_da)-\sqrt{\varepsilon_r}J_n(ka)J_n^{\prime}(k_da)}{\sqrt{\mu_r}H_n^{(2)\prime}(ka)J_n(k_da)-\sqrt{\varepsilon_r}H_n^{(2)}(ka)J_n^{\prime}(k_da)}
		\end{aligned}
	\end{equation}
	where $(\rho^{\prime}, \phi^{\prime})$ is the coordinates of the line source, $k$ and $k_d$ are the wavenumber in free space and dielectric cylinder, respectively. The SNR distributions of interpolation results for different frequencies are shown in Fig. \ref{Fig4}. The contrast of DoI ranges from 0 to 5 with the step of 0.1, so there are 50 samples for each frequency. From the figure, we can see that the SNR distribution of $N_{\text{DoF}}^{(3)}$ is the highest, followed by $N_{\text{DoF}}^{(2)}$, $N_{\text{DoF}}^{(1)}$, and $N_{\text{DoF}}^{(4)}$. The mean SNR of interpolated fields for $N_{\text{DoF}}^{(1)}$, $N_{\text{DoF}}^{(2)}$, $N_{\text{DoF}}^{(3)}$, and $N_{\text{DoF}}^{(4)}$ are 35.7 dB, 60.6 dB, 66.7 dB and 14.3 dB, respectively. We plot the interpolation results of one example in Fig. \ref{Fig5}. The contrast of the cylinder is 5, and the operating frequency is 5 GHz. It can be seen that the interpolated fields of $N_{\text{DoF}}^{(2)}$ and $N_{\text{DoF}}^{(3)}$ are basically the same as the ground truth, and that of $N_{\text{DoF}}^{(1)}$ is slightly different in amplitude from the ground truth, but that of $N_{\text{DoF}}^{(4)}$ rarely agrees with the ground truth. Therefore, $N_{\text{DoF}}^{(4)}$ field samples are not enough in measurement, especially for strong scatterers. On the other hand, $N_{\text{DoF}}^{(1)}$ meets the basic requirement for the NDoF of scattered field, while $N_{\text{DoF}}^{(2)}$ and $N_{\text{DoF}}^{(3)}$ provide an advisable estimation for the NDoF with high accuracy.
	
	\subsection{Random Dielectric Samples}
	
	\begin{figure}[!tbp]
		\centering
		\includegraphics[width=\linewidth]{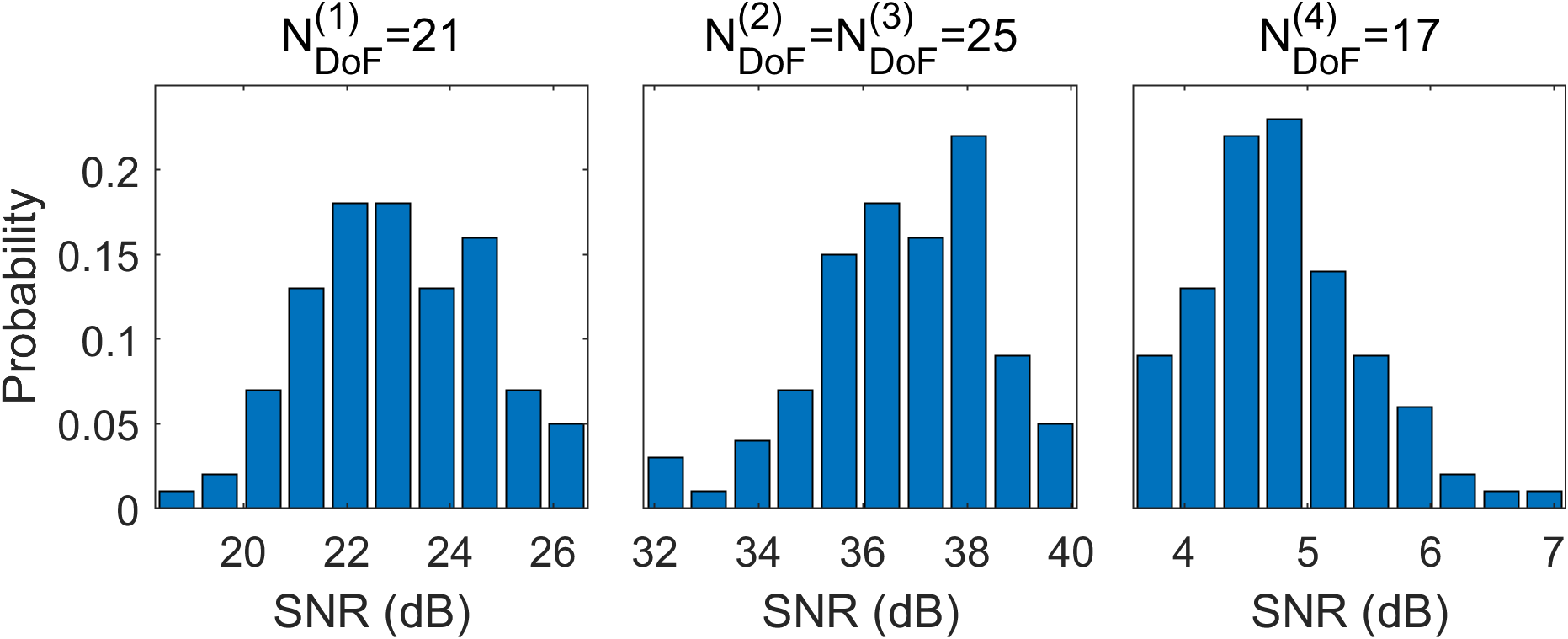}
		\caption{SNR distribution of interpolation results of 100 random samples with circular DoI.}
		\label{Fig7}
	\end{figure}

	\begin{figure}[!tbp]
	\centering
	\subfigure{
		\includegraphics[width=0.45\linewidth]{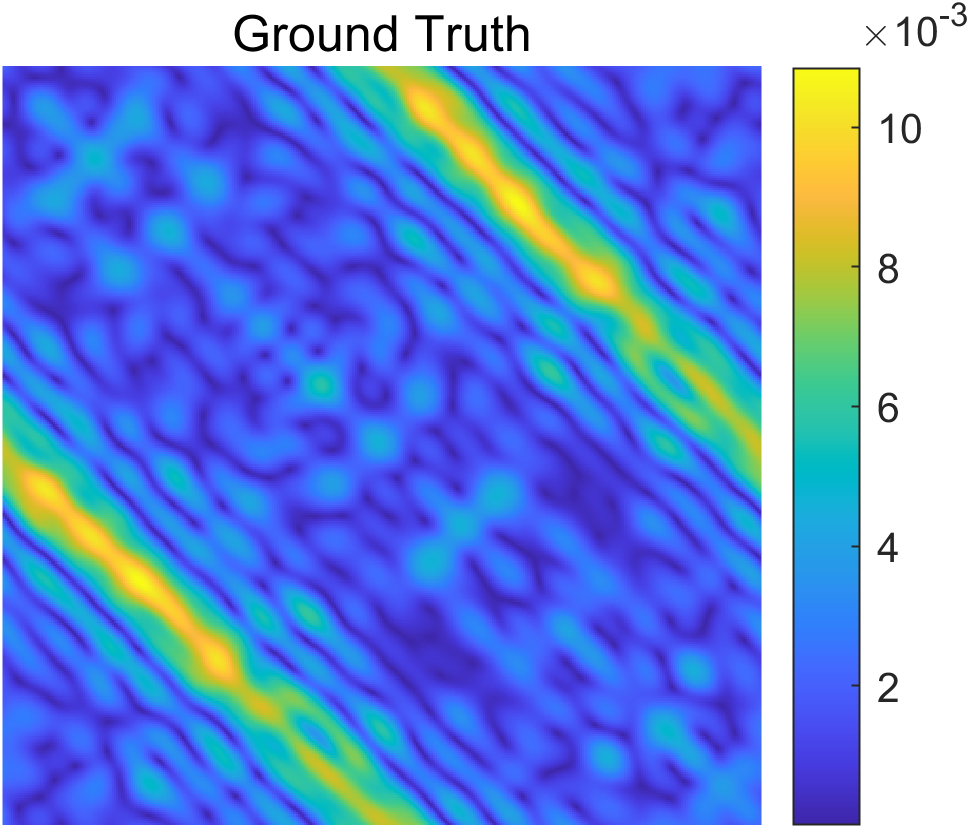}\label{Fig8a}} \\
	\subfigure{
		\includegraphics[width=0.9\linewidth]{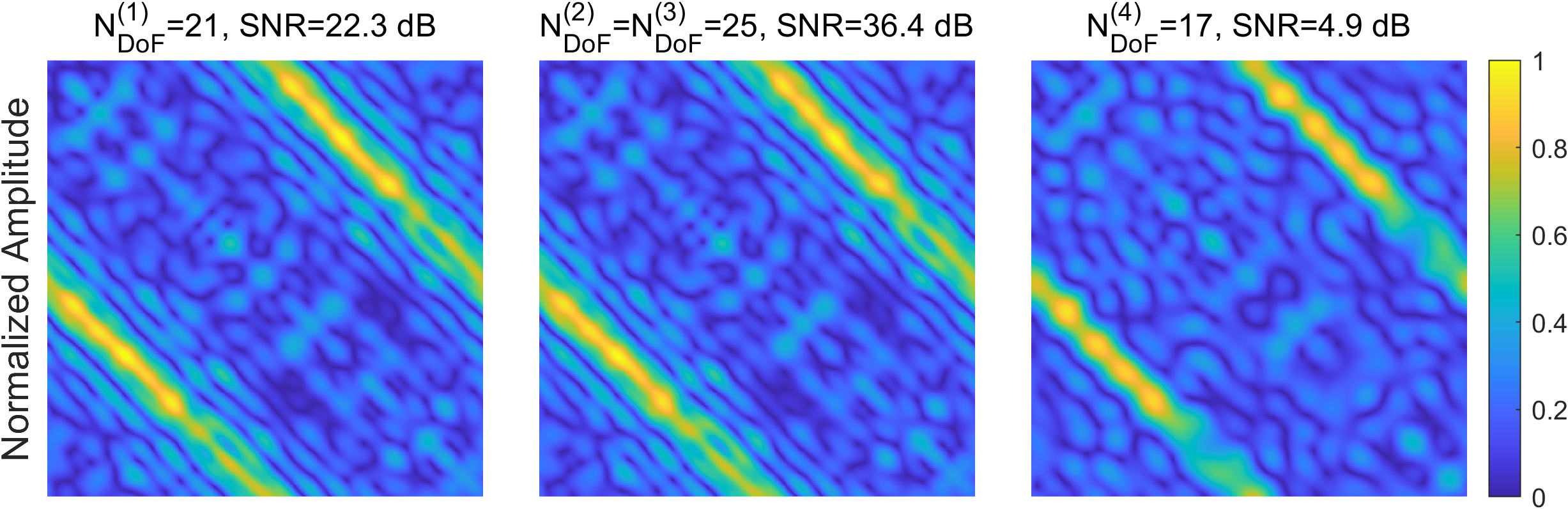}\label{Fig8b}}
	\subfigure{
		\includegraphics[width=0.9\linewidth]{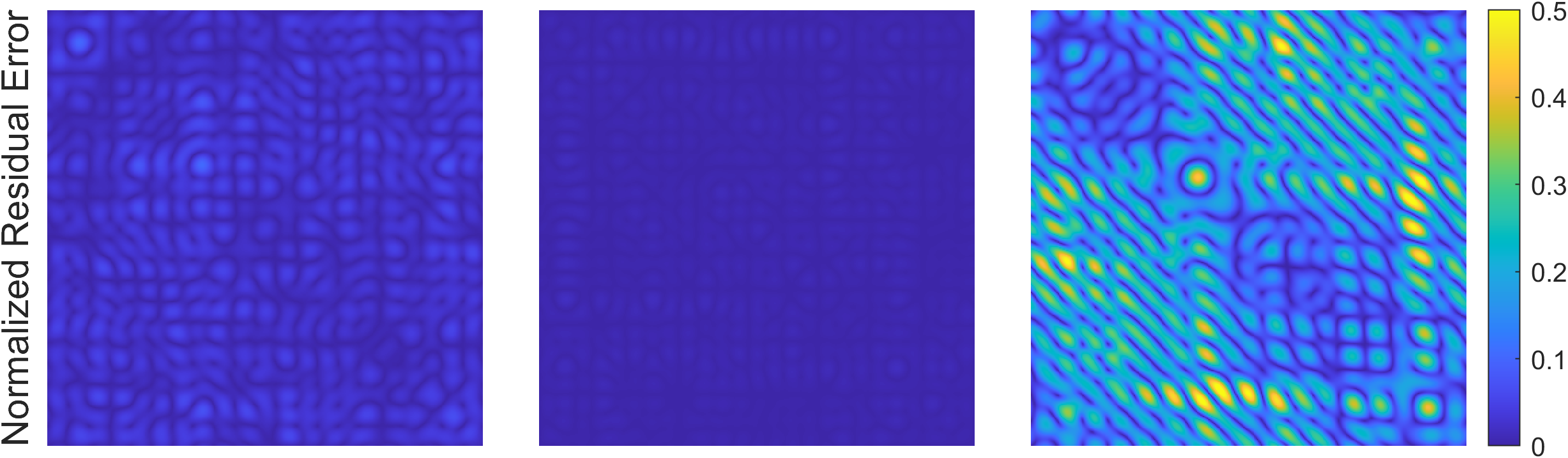}\label{Fig8c}} \\
	\caption{The interpolated scattered fields of one random example with circular DoI. The filed amplitude of the second and third row are normalized by the maximum field amplitude of the ground truth.}
	\label{Fig8}
\end{figure}
	
	In this subsection, we verify the proposed NDoF with random dielectric samples. Firstly, 100 test samples with circular DoI are generated, where the contrast of each subdomain is randomly sampled from 0 to 5. One of the test samples is shown in Fig. \ref{Fig6a}. We set the operating frequency to 4 GHz and $\chi$ to 5, thus $N_{\text{DoF}}^{(1)}=21$,  $N_{\text{DoF}}^{(2)}=N_{\text{DoF}}^{(3)}=25$, and $N_{\text{DoF}}^{(4)}=17$. Again, we interpolate the field samples of size $N_{\text{DoF}}\times N_{\text{DoF}}$ to the oversampled fields with the frequency-domain zero-padding method. The SNR distributions of interpolation results of four NDoF are summarized in Fig. \ref{Fig7}, where the mean values of the three distributions are 23.0 dB, 36.7 dB, and 4.8 dB, respectively. 
	The interpolation results of Fig. \ref{Fig6a} are illustrated in Fig. \ref{Fig8}. The first row is the field amplitude of the ground truth. The second and third rows are the normalized amplitude of the interpolation results and the residual error between the interpolation results and the ground truth, they are all normalized by the maximum field amplitude of the ground truth. From the figure, we can see that $N_{\text{DoF}}^{(3)}\times N_{\text{DoF}}^{(3)}$ field samples are enough to capture the information of the oversampled scattered fields, thus verifying the accuracy of the proposed methods in calculating the NDoF. 
	
	\begin{figure}[!tbp]
		\centering
		\includegraphics[width=\linewidth]{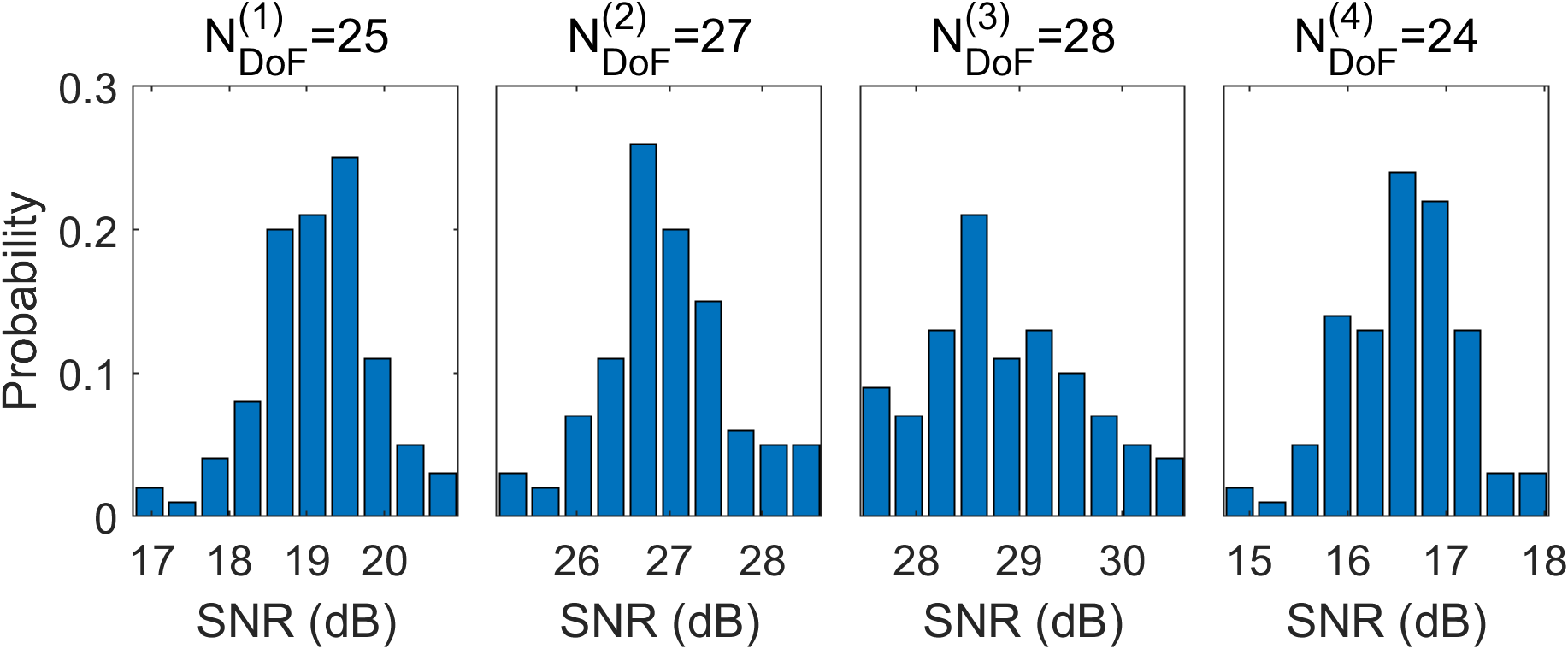}
		\caption{SNR distribution of interpolation results of 100 random samples with square DoI.}
		\label{Fig9}
	\end{figure}

	\begin{figure}[!tbp]
	\centering
	\subfigure{
		\includegraphics[width=0.45\linewidth]{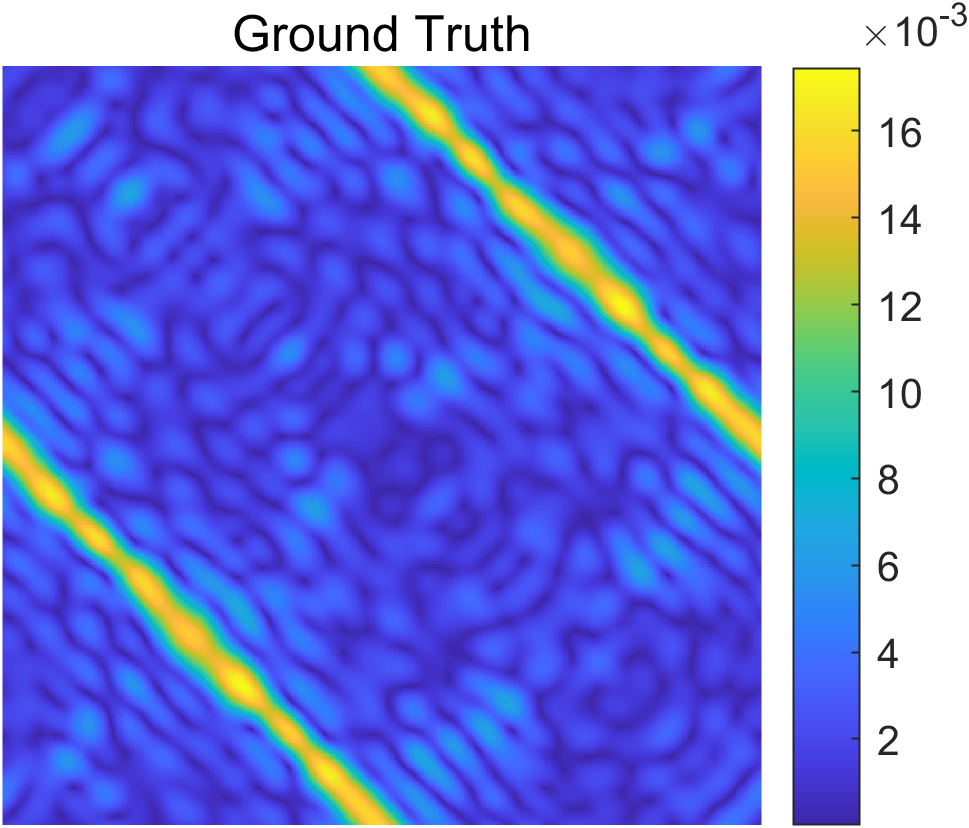}\label{Fig10a}} \\
	\subfigure{
		\includegraphics[width=\linewidth]{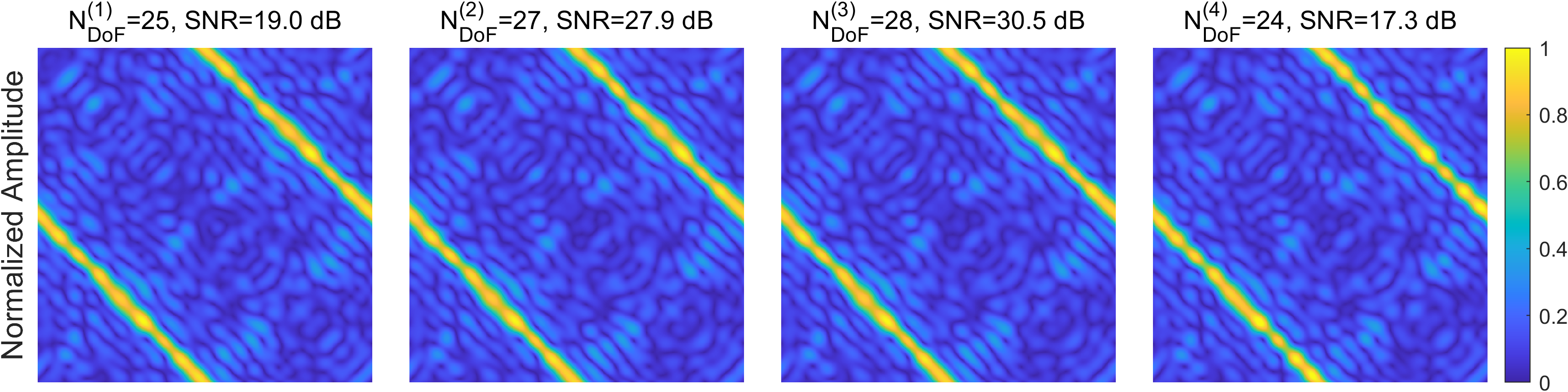}\label{Fig10b}}
	\subfigure{
		\includegraphics[width=\linewidth]{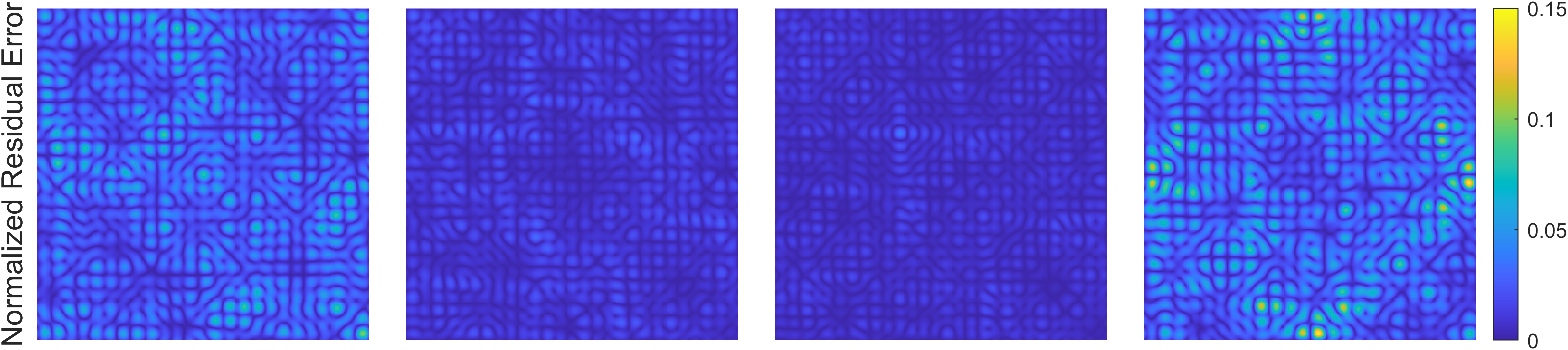}\label{Fig10c}} \\
	\caption{The interpolated scattered fields of one random example with square DoI. The filed amplitude of the second and third row are normalized by the maximum field amplitude of the ground truth.}
	\label{Fig10}
	\end{figure}
	
	In order to keep the properties of the singular values of the discretized radiation matrix consistent with those of the analytical radiation operator, all the above analyses are performed on a circular DoI, where $\bar{\bar{G}}_D$ and $\bar{\bar{G}}_S$ are computed based on the 12892 subdomains. However, since the commonly used DoI in numerical calculations is square, we also study the accuracy of the proposed NDoF on square DoI. Here, we generate 100 random samples with square DoI, as shown in Fig. \ref{Fig6b}. Since the DoI is square, the definition of $N_{\text{DoF}}^{(4)}$ is modified as the smallest integer larger than $2\sqrt{2}\beta a$, and  $\bar{\bar{G}}_D$ and $\bar{\bar{G}}_S$ are computed based on the 128$\times$128 subdomains. The experiment setups are the same as the former one, thus $N_{\text{DoF}}^{(1)}=25$,  $N_{\text{DoF}}^{(2)}=27, N_{\text{DoF}}^{(3)}=28$, and $N_{\text{DoF}}^{(4)}=24$. They are all slightly increased compared to the circular DoI, which is reasonable since the DoI is larger. The SNR distributions of interpolation results for four NDoF are summarized in Fig. \ref{Fig9}, where the mean values of these distributions are 19.1 dB, 27.0 dB, 28.8 dB, and 16.6 dB, respectively. The interpolation results of Fig. \ref{Fig6b} are illustrated in Fig. \ref{Fig10}. We can see that the interpolation results of $N_{\text{DoF}}^{(2)}$ and $N_{\text{DoF}}^{(3)}$ are still satisfactory. It is noted that for the square DoI, we are supposed to compute the radiation matrices based on its smallest circumscribed circle in a strict sense. Nevertheless, the above experiment shows that the reduced accuracy of the proposed NDoF is acceptable even though the radiation matrices are computed based on the square DoI.
	
	\subsection{Experimental Example}
	\begin{figure}[!tbp]
		\centering
		\includegraphics[width=0.45\linewidth]{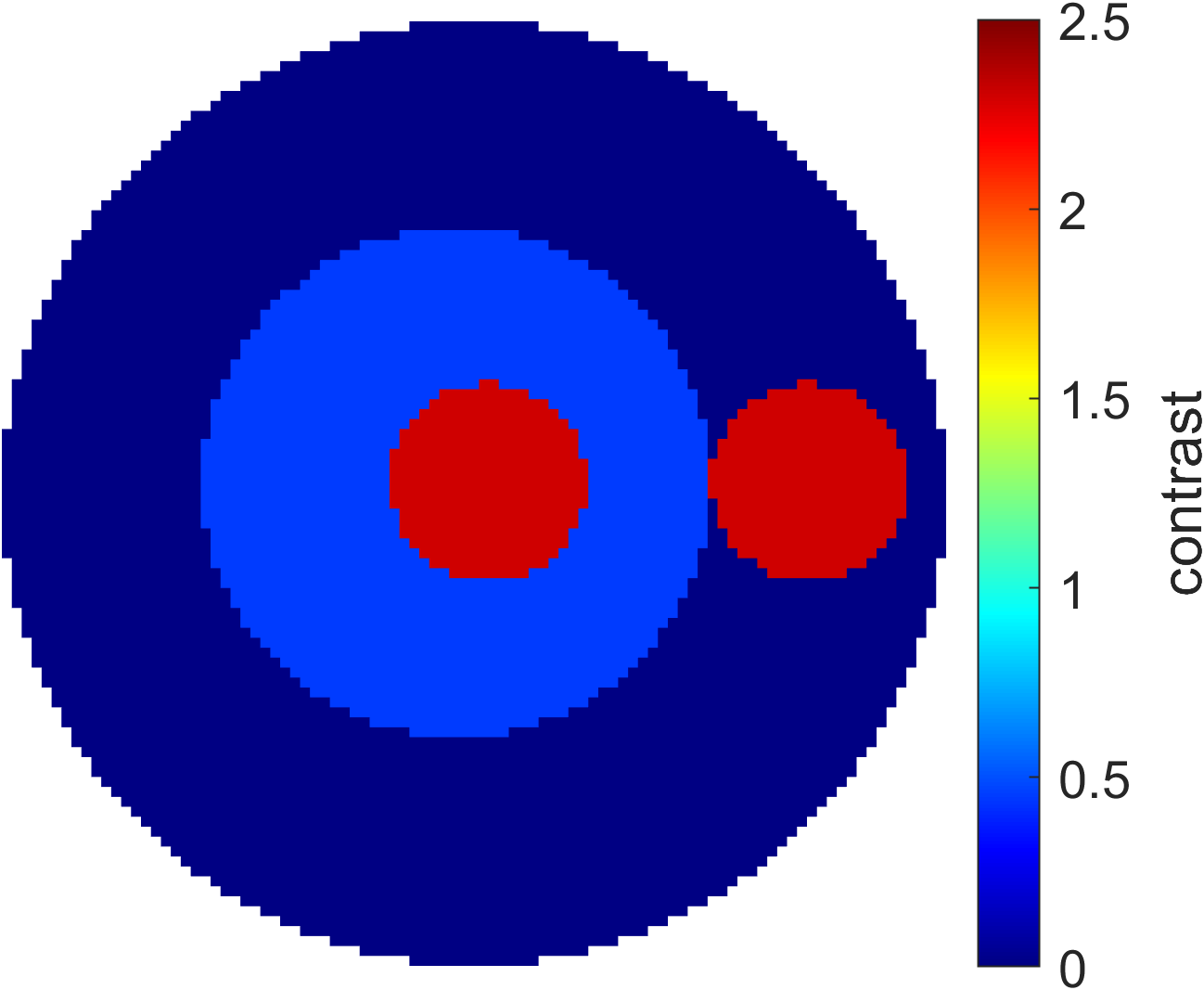}
		\caption{Schematic diagram of \textit{FoamTwinDiel} model. The contrast of two types of cylinders are 2.3 and 1.45, respectively. The radius of the DoI is 0.075 m.}
		\label{Fig11}
	\end{figure}
	
	\begin{figure}[!tbp]
		\centering
		\includegraphics[width=0.95\linewidth]{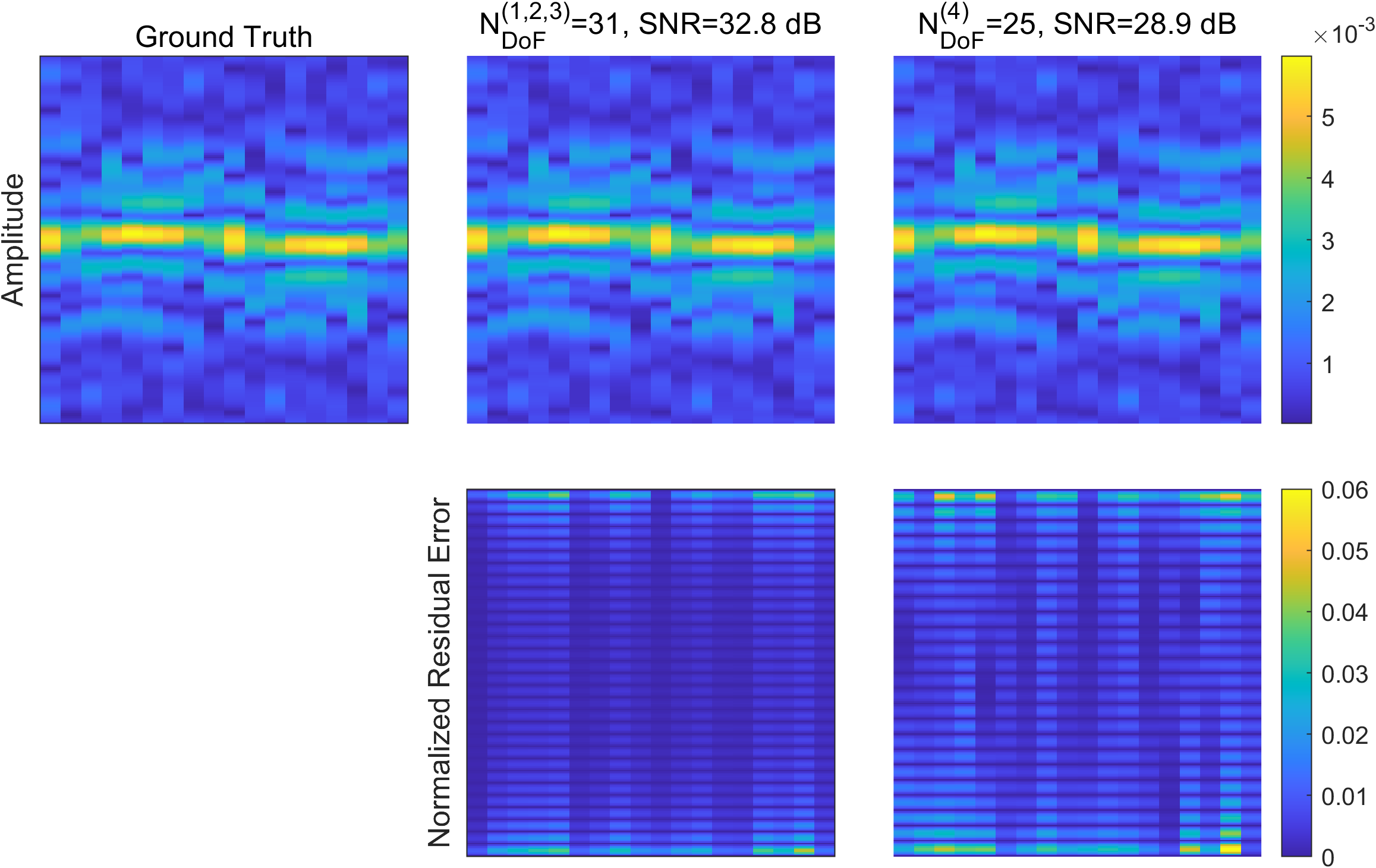}
		\caption{The interpolated scattered fields of \textit{FoamTwinDiel} model. The residual errors are normalized by the maximum amplitude of the ground truth.}
		\label{Fig12}
	\end{figure}

	In the last experiment, we test the proposed NDoF with experimental data from Institut Fresnel \cite{geffrin2005free}. The scattered field of \textit{FoamTwinDiel} model is employed. This model consists of two types of dielectric scatterers, including two small cylinders with contrast of 2.3 and diameter of 31 mm, and a large cylinder with contrast of 1.45 and diameter of 80 mm, as shown in Fig. \ref{Fig11}. The DoI for computing the radiation matrices of this model is set as a circle with radius of 75 mm in 2D geometry. The numbers of transmitters and receivers are 18 and 241, respectively. It is noted that the range of observation angle for each illumination is only $241^\circ$ out of $360^\circ$. More details of the measurement setup can be found in \cite{geffrin2005free}. Since the field samples are only available in the range of 241 degrees, the traditional NDoF $2\beta a$ that is for the full-view field should be scaled down to $241/360$ of the original. Here we ignore the spectral leakage caused by the truncation effect. Thus, we modify $N_{\text{DoF}}^{(4)}$ as the smallest integer larger than $2\beta a\times 241/360$. The operating frequency is set to 10 GHz, and $\chi$ is set to 2.3, thus we can calculate that $N_{\text{DoF}}^{(1)}=26$,  $N_{\text{DoF}}^{(2)}=29, N_{\text{DoF}}^{(3)}=29$, and $N_{\text{DoF}}^{(4)}=22$. It is noted that $\bar{\bar{G}}_S$ of each transmitter is different due to the measurement setup, but the calculated $N_{\text{DoF}}^{(i)} (i=1,2,3)$ are the same. Besides, since the scattered field is not periodic, the frequency-domain zero-padding method is no longer the optimal interpolation algorithm in this case, but the increase of interpolation error is tolerable. To avoid the influence of extrapolation, we restrict the sampling interval to the factors of 240. Then, $N_{\text{DoF}}^{(1)} = N_{\text{DoF}}^{(2)} = N_{\text{DoF}}^{(3)}=31, N_{\text{DoF}}^{(4)}=25$. In this experiment, we only interpolate the field samples in the receiver dimension. The interpolation results are summarized in Fig. \ref{Fig12}, where the ground truth is the experimental data. We can see that despite the measurement errors, the interpolated scattered fields of $N_{\text{DoF}}^{(i)} (i=1,2,3)$ are basically the same as the ground truth. Therefore, 31 field samples are enough to include the information of the 241 field samples.
	
	\section{Discussion} \label{sec5}
	The proposed definition of the NDoF associates the spatial bandwidth of the scattered field with the number of principal singular values of the radiation matrix. This definition is deduced from the analytical analyses of the radiation operator performed on a circular DoI \cite{bucci1999improving}. In that case, the rationality of the definition of NDoF can be explained as follows. For the Fourier transform, the scattered field can be regarded as being expanded with plane waves. For singular system of the circular DoI, the scattered field is expanded with cylindrical waves. Since these two basis functions can be converted to each other, the number of principal singular values is basically equal to the spatial bandwidth of the scattered field.
	
	The predefined fraction $P$ that controls the approximation error of scattered field data is obtained from numerical experiments. For a more general case, $P$ can be determined according to $N_{\text{DoF}}^{(1)}$. Specifically, since the singular values of $\bar{\bar{G}}_S$ behave in the same way as analytical singular values, i.e., $\hat{\sigma}_{n}$ in \eqref{eq3}, we can adjust $P$ in $N_{\text{DoF}}^{(1)}$ to find the corresponding $P$ when $N_{\text{DoF}}^{(1)} \approx 2\beta a$. Here $2\beta a$ is used as the benchmark for estimating NDoF. This $P$ is then applied to calculate $N_{\text{DoF}}^{(2)}$ and $N_{\text{DoF}}^{(3)}$. In fact, $P$ should be proportional to $\beta a$ considering that the knee is steeper when $\beta a$ is larger, as shown in Fig. \ref{Fig3b}. But for convenience, we consider $P=99\%$ as a typical value, which can cover most cases for $\beta a<30$ according to our numerical experiments. Besides, we observe from inversion experiments that $N_{\text{DoF}}^{(2)}(P=99\%)$ is a good estimate for the threshold, above which the image quality of inversion result no longer improves significantly. Due to page limitation, the inversion experiments are not presented here.
	
	For any scatterers, we are supposed to compute the radiation matrices $\bar{\bar{G}}_D$ and $\bar{\bar{G}}_S$ based on the smallest circle enclosing the scatterers in 2D geometry to obtain a relatively accurate NDoF. For simplicity, numerical experiments show that the proposed NDoF can be estimated by the minimum integer larger than $2\beta a$ plus a proper constant, where $a$ is the radius of the smallest circle surrounding the scatterers. In most cases of inverse scattering study where $\beta a<30$, the approximate upper bound $N_{\text{DoF}}^{(3)}$ can be estimated by $\lceil 2\beta a \rceil + 8$. Apart from the circular DoI, numerical examples also show that it is feasible to compute the radiation matrices based on the square DoI enclosing the scatterers to obtain the proposed NDoF. However, the accuracy may drop within an acceptable range.
	
	\section{Conclusion} \label{sec6}
	We study the impact of scatterer contrast on the NDoF of scattered field through numerical experiments, which could be useful for nonlinear ISP, especially for the inversions of large contrast targets. Three numerical methods for calculating the NDoF of scattered field in nonlinear ISP are presented. The properties of the singular values of numerical radiation matrices are studied. The first NDoF is defined as the number of principle singular values of the external radiation matrix, which is used as a benchmark. To take into account the contrast of DoI for calculating the second NDoF, we modify the radiation matrix under the assumption that the DoI is homogeneous. Furthermore, after exploring the relationship between NDoF and the maximum contrast of DoI, the approximate upper bound of the second NDoF is obtained, that is, the third NDoF. It is noted that the third NDoF is independent of the contrast. The accuracy of the proposed NDoF is verified by interpolating field samples of NDoF to oversampled filed. Interpolation experiments using both synthetic and experimental data validate the effectiveness of the proposed NDoF. 
	
	One of the next steps would be to investigate analytically the phenomena exhibited by the above numerical experiments. The effect of scatterer contrast on the second NDoF may need a physical analysis. The relationship between the second NDoF and the third NDoF could be discussed mathematically. In addition, the reason that the third NDoF can be estimated by $\lceil 2\beta a \rceil + 8$ for $\beta a<30$ could be further studied. Moreover, this numerical analysis method for NDoF can be  extended to three-dimensional inverse scattering scenes. This will be our future work.
	
	\bibliographystyle{IEEEtran}
	\bibliography{IEEEexample_v2} 

\begin{thebibliography}{10}
\providecommand{\url}[1]{#1}
\csname url@samestyle\endcsname
\providecommand{\newblock}{\relax}
\providecommand{\bibinfo}[2]{#2}
\providecommand{\BIBentrySTDinterwordspacing}{\spaceskip=0pt\relax}
\providecommand{\BIBentryALTinterwordstretchfactor}{4}
\providecommand{\BIBentryALTinterwordspacing}{\spaceskip=\fontdimen2\font plus
\BIBentryALTinterwordstretchfactor\fontdimen3\font minus
  \fontdimen4\font\relax}
\providecommand{\BIBforeignlanguage}[2]{{%
\expandafter\ifx\csname l@#1\endcsname\relax
\typeout{** WARNING: IEEEtran.bst: No hyphenation pattern has been}%
\typeout{** loaded for the language `#1'. Using the pattern for}%
\typeout{** the default language instead.}%
\else
\language=\csname l@#1\endcsname
\fi
#2}}
\providecommand{\BIBdecl}{\relax}
\BIBdecl

\bibitem{chen2018computational}
X.~Chen, \emph{Computational methods for electromagnetic inverse
  scattering}.\hskip 1em plus 0.5em minus 0.4em\relax Wiley Online Library,
  2018.

\bibitem{woodhouse2017introduction}
I.~H. Woodhouse, \emph{Introduction to Microwave Remote Sensing}.\hskip 1em
  plus 0.5em minus 0.4em\relax CRC press, 2017.

\bibitem{9325547}
M.~{Li}, R.~{Guo}, K.~{Zhang}, Z.~{Lin}, F.~{Yang}, S.~{Xu}, X.~{Chen},
  A.~{Massa}, and A.~{Abubakar}, ``Machine learning in electromagnetics with
  applications to biomedical imaging: A review,'' \emph{IEEE Antennas and
  Propagation Magazine}, 2021.

\bibitem{abubakar20082}
A.~Abubakar, T.~Habashy, V.~Druskin, L.~Knizhnerman, and D.~Alumbaugh, ``2.5{D}
  forward and inverse modeling for interpreting low-frequency electromagnetic
  measurements,'' \emph{Geophysics}, vol.~73, no.~4, pp. F165--F177, 2008.

\bibitem{salucci2016real}
M.~Salucci, N.~Anselmi, G.~Oliveri, P.~Calmon, R.~Miorelli, C.~Reboud, and
  A.~Massa, ``Real-time {NDT-NDE} through an innovative adaptive partial least
  squares {SVR} inversion approach,'' \emph{IEEE Transactions on Geoscience and
  Remote Sensing}, vol.~54, no.~11, pp. 6818--6832, 2016.

\bibitem{bucci1987spatial}
O.~Bucci and G.~Franceschetti, ``On the spatial bandwidth of scattered
  fields,'' \emph{IEEE Transactions on Antennas and Propagation}, vol.~35,
  no.~12, pp. 1445--1455, 1987.

\bibitem{bucci1989degrees}
O.~M. Bucci and G.~Franceschetti, ``On the degrees of freedom of scattered
  fields,'' \emph{IEEE Transactions on Antennas and Propagation}, vol.~37,
  no.~7, pp. 918--926, 1989.

\bibitem{bucci1997electromagnetic}
O.~Bucci and T.~Isernia, ``Electromagnetic inverse scattering: Retrievable
  information and measurement strategies,'' \emph{Radio Science}, vol.~32,
  no.~6, pp. 2123--2137, 1997.

\bibitem{bucci1999improving}
O.~Bucci, L.~Crocco, and T.~Isernia, ``Improving the reconstruction
  capabilities in inverse scattering problems by exploitation of
  close-proximity setups,'' \emph{JOSA A}, vol.~16, no.~7, pp. 1788--1798,
  1999.

\bibitem{bucci1991optimal}
O.~M. Bucci, C.~Gennarelli, and C.~Savarese, ``Optimal interpolation of
  radiated fields over a sphere,'' \emph{IEEE Transactions on Antennas and
  Propagation}, vol.~39, no.~11, pp. 1633--1643, 1991.

\bibitem{bucci1993interpolation}
O.~Bucci, C.~Gennarelli, and C.~Savarese, ``Interpolation of electromagnetic
  radiated fields over a plane by nonuniform samples,'' \emph{IEEE Transactions
  on Antennas and Propagation}, vol.~41, no.~11, pp. 1501--1508, 1993.

\bibitem{bucci1998representation}
O.~M. Bucci, C.~Gennarelli, and C.~Savarese, ``Representation of
  electromagnetic fields over arbitrary surfaces by a finite and nonredundant
  number of samples,'' \emph{IEEE Transactions on Antennas and Propagation},
  vol.~46, no.~3, pp. 351--359, 1998.

\bibitem{brancaccio1998information}
A.~Brancaccio, G.~Leone, and R.~Pierri, ``Information content of born scattered
  fields: results in the circular cylindrical case,'' \emph{JOSA A}, vol.~15,
  no.~7, pp. 1909--1917, 1998.

\bibitem{pierri1999information}
R.~Pierri, R.~Persico, and R.~Bernini, ``Information content of the born field
  scattered by an embedded slab: multifrequency, multiview, and
  multifrequency--multiview cases,'' \emph{JOSA A}, vol.~16, no.~10, pp.
  2392--2399, 1999.

\bibitem{di1969degrees}
G.~T. Di~Francia, ``Degrees of freedom of an image,'' \emph{JOSA}, vol.~59,
  no.~7, pp. 799--804, 1969.

\bibitem{piestun2000electromagnetic}
R.~Piestun and D.~A. Miller, ``Electromagnetic degrees of freedom of an optical
  system,'' \emph{JOSA A}, vol.~17, no.~5, pp. 892--902, 2000.

\bibitem{bertero1989linear}
M.~Bertero, ``Linear inverse and ill-posed problems,'' \emph{Advances in
  Electronics and Electron Physics}, vol.~75, pp. 1--120, 1989.

\bibitem{pierri1998information}
R.~Pierri and F.~Soldovieri, ``On the information content of the radiated
  fields in the near zone over bounded domains,'' \emph{Inverse Problems},
  vol.~14, no.~2, p. 321, 1998.

\bibitem{pierri2020asymptotic}
R.~Pierri and R.~Moretta, ``Asymptotic study of the radiation operator for the
  strip current in near zone,'' \emph{Electronics}, vol.~9, no.~6, p. 911,
  2020.

\bibitem{pierri2021ndf}
------, ``{NDF} of the near-zone field on a line perpendicular to the source,''
  \emph{IEEE Access}, vol.~9, pp. 91\,649--91\,660, 2021.

\bibitem{leone2022dimension}
G.~Leone, R.~Moretta, and R.~Pierri, ``Dimension and sampling of the near-field
  and its intensity over curves,'' \emph{IEEE Open Journal of Antennas and
  Propagation}, 2022.

\bibitem{pierri2021svd}
R.~Pierri and R.~Moretta, ``An {SVD} approach for estimating the dimension of
  phaseless data on multiple arcs in fresnel zone,'' \emph{Electronics},
  vol.~10, no.~5, p. 606, 2021.

\bibitem{pierri2021dimension}
R.~Pierri, G.~Leone, and R.~Moretta, ``The dimension of phaseless near-field
  data by asymptotic investigation of the lifting operator,''
  \emph{Electronics}, vol.~10, no.~14, p. 1658, 2021.

\bibitem{moretta2022optimal}
R.~Moretta, G.~Leone, F.~Munno, and R.~Pierri, ``Optimal field sampling of arc
  sources via asymptotic study of the radiation operator,'' \emph{Electronics},
  vol.~11, no.~2, p. 270, 2022.

\bibitem{migliore2020cares}
M.~D. Migliore, ``Who cares about the horse? a gentle introduction to
  information in electromagnetic theory [wireless corner],'' \emph{IEEE
  Antennas and Propagation Magazine}, vol.~62, no.~5, pp. 126--137, 2020.

\bibitem{migliore2008electromagnetics}
------, ``On electromagnetics and information theory,'' \emph{IEEE Transactions
  on Antennas and Propagation}, vol.~56, no.~10, pp. 3188--3200, 2008.

\bibitem{loyka2018information}
S.~Loyka and J.~Mosig, ``Information theory and electromagnetism: Are they
  related?'' in \emph{MIMO System Technology for Wireless
  Communications}.\hskip 1em plus 0.5em minus 0.4em\relax CRC Press, 2018, pp.
  57--88.

\bibitem{gruber2008new}
F.~K. Gruber and E.~A. Marengo, ``New aspects of electromagnetic information
  theory for wireless and antenna systems,'' \emph{IEEE Transactions on
  Antennas and Propagation}, vol.~56, no.~11, pp. 3470--3484, 2008.

\bibitem{franceschetti2015information}
M.~Franceschetti, M.~D. Migliore, P.~Minero, and F.~Schettino, ``The
  information carried by scattered waves: Near-field and nonasymptotic
  regimes,'' \emph{IEEE Transactions on Antennas and Propagation}, vol.~63,
  no.~7, pp. 3144--3157, 2015.

\bibitem{poon2005degrees}
A.~S. Poon, R.~W. Brodersen, and D.~N. Tse, ``Degrees of freedom in
  multiple-antenna channels: A signal space approach,'' \emph{IEEE Transactions
  on Information Theory}, vol.~51, no.~2, pp. 523--536, 2005.

\bibitem{hanlen2006wireless}
L.~Hanlen and M.~Fu, ``Wireless communication systems with-spatial diversity: A
  volumetric model,'' \emph{IEEE Transactions on Wireless Communications},
  vol.~5, no.~1, pp. 133--142, 2006.

\bibitem{xu2006electromagnetic}
J.~Xu and R.~Janaswamy, ``Electromagnetic degrees of freedom in 2-d scattering
  environments,'' \emph{IEEE Transactions on Antennas and Propagation},
  vol.~54, no.~12, pp. 3882--3894, 2006.

\bibitem{janaswamy2011degrees}
R.~Janaswamy, ``On the {EM} degrees of freedom in scattering environments,''
  \emph{IEEE Transactions on Antennas and Propagation}, vol.~59, no.~10, pp.
  3872--3881, 2011.

\bibitem{dickey2003super}
F.~Dickey, L.~Romero, J.~DeLaurentis, and A.~Doerry, ``Super-resolution,
  degrees of freedom and synthetic aperture radar,'' \emph{IEE
  Proceedings-Radar, Sonar and Navigation}, vol. 150, no.~6, pp. 419--429,
  2003.

\bibitem{krishnan2011synthetic}
V.~Krishnan and B.~Yazici, ``Synthetic aperture radar imaging exploiting
  multiple scattering,'' \emph{Inverse Problems}, vol.~27, no.~5, p. 055004,
  2011.

\bibitem{solimene2014sar}
R.~Solimene, I.~Catapano, G.~Gennarelli, A.~Cuccaro, A.~Dell'Aversano, and
  F.~Soldovieri, ``{SAR} imaging algorithms and some unconventional
  applications: A unified mathematical overview,'' \emph{IEEE Signal Processing
  Magazine}, vol.~31, no.~4, pp. 90--98, 2014.

\bibitem{mehrotra2020dof}
N.~Mehrotra and A.~Sabharwal, ``{DoF} analysis for multipath-assisted imaging:
  Single frequency illumination,'' in \emph{2020 IEEE International Symposium
  on Information Theory (ISIT)}.\hskip 1em plus 0.5em minus 0.4em\relax IEEE,
  2020, pp. 1456--1461.

\bibitem{lyons2010interpolate}
R.~Lyons, ``How to interpolate in the time-domain by zero-padding in the
  frequency domain,'' \emph{H{\"a}mtat fr{\aa}n dspGuru by Iowegian
  International Corporation}, 2010.

\bibitem{lyons2010understanding}
R.~G. Lyons, \emph{Understanding Digital Signal Processing: Unders Digita
  Signal Proces\_3}.\hskip 1em plus 0.5em minus 0.4em\relax Pearson Education,
  2010.

\bibitem{jin2011theory}
J.-M. Jin, \emph{Theory and computation of electromagnetic fields}.\hskip 1em
  plus 0.5em minus 0.4em\relax John Wiley \& Sons, 2011.

\bibitem{chen2009subspace}
X.~Chen, ``Subspace-based optimization method for solving inverse-scattering
  problems,'' \emph{IEEE Transactions on Geoscience and Remote Sensing},
  vol.~48, no.~1, pp. 42--49, 2009.

\bibitem{zhong2009twofold}
Y.~Zhong and X.~Chen, ``Twofold subspace-based optimization method for solving
  inverse scattering problems,'' \emph{Inverse Problems}, vol.~25, no.~8, p.
  085003, 2009.

\bibitem{lin2021low}
Z.~Lin, R.~Guo, M.~Li, A.~Abubakar, T.~Zhao, F.~Yang, and S.~Xu,
  ``Low-frequency data prediction with iterative learning for highly nonlinear
  inverse scattering problems,'' \emph{IEEE Transactions on Microwave Theory
  and Techniques}, 2021.

\bibitem{geffrin2005free}
J.-M. Geffrin, P.~Sabouroux, and C.~Eyraud, ``Free space experimental
  scattering database continuation: experimental set-up and measurement
  precision,'' \emph{Inverse Problems}, vol.~21, no.~6, p. S117, 2005.

\end{thebibliography}
\end{document}